\documentclass[a4paper,11pt]{article} \usepackage[margin=1.4in]{geometry}

\usepackage{dsfont,afterpage}

\usepackage[normalem]{ulem}
\usepackage{afterpage}
\usepackage{xurl}
\usepackage{hyperref}
\hypersetup{
 colorlinks  = true, %Colours links instead of ugly boxes
 urlcolor   = black, %Colour for external hyperlinks
 linkcolor  = black, %Colour of internal links
 citecolor  = black %Colour of citations
}

\makeatletter
\renewcommand\@makefnmark{\hbox{\@textsuperscript{\normalfont\color{blue}\@thefnmark}}}
\renewcommand\@makefntext[1]{%
 \parindent 1em\noindent
      \hb@xt@1.8em{%
        \hss\@textsuperscript{\normalfont\@thefnmark}}#1}
\makeatother

\usepackage[hang,flushmargin]{footmisc}
\makeatletter
\newcommand{\algorithmfootnote}[2][\footnotesize]{%
  \let\old@algocf@finish\@algocf@finish% Store algorithm finish macro
  \def\@algocf@finish{\old@algocf@finish% Update finish macro to insert "footnote"
    \leavevmode\rlap{\begin{minipage}{\linewidth}
    #1#2\vspace*{-7pt}

    \end{minipage}}%\vspace*{-1cm}

  }%
}
\makeatother
\makeatletter
\def\blfootnote{\gdef\@thefnmark{}\@footnotetext}
\makeatother
\usepackage{rotating}

\usepackage{natbib} 
\def\citeN{\citet*}

\usepackage{bbm}
\usepackage{adjustbox}
\usepackage[linesnumbered,vlined,boxed,commentsnumbered]{algorithm2e}
\usepackage{amsthm}
\usepackage[table,dvipsnames]{xcolor}
\usepackage{xspace}
\usepackage{diagbox}

\usepackage{glossaries}
\usepackage{lmodern}            % Latin Modern
\usepackage[T1]{fontenc}        % Font encription
\usepackage[utf8]{inputenc}    % input encription utf-8

\usepackage{multirow}
\usepackage{amssymb} 
\usepackage{amsmath}

\usepackage{afterpage}
\usepackage{placeins} % FloatBarrier

\usepackage{hhline}
\usepackage{arydshln}
\usepackage{caption}
\usepackage{subcaption}
\usepackage{float}
\usepackage{graphicx}
\usepackage{lscape}

\DeclareMathOperator*{\argmin}{arg\,min}

\def\LGD{\mathcal{C}\xspace}\def\LGD{{\rm LGD}\xspace}
\def\CVA{{\rm CVA}\xspace}
\def\E{\mathbb{E}} 
\def\ES{\mathbb{ES}}

\def\strike{c}\def\strike{k}
\renewcommand\b[1]{ {#1} } 
\renewcommand\b[1]{ {\color{blue}#1} }
\renewcommand\r[1]{} \renewcommand\r[1]{ {\color{red}#1} }
\def\sp{,\;}

\newcommand{\beqa}{\begin{eqnarray}}
\newcommand{\eeqa}{\end{eqnarray}}
\newcommand{\bea}{\begin{eqnarray*}}
\newcommand{\eea}{\end{eqnarray*}}
\def\bal{\begin{aligned}}
\def\eal{\end{aligned}}
\def\bll#1{\beqa\label{#1}\bal}
\def\lel{\eal\eeqa}
\newcommand{\beql}[1]{\bll{#1}}
\newcommand{\eeql}{\lel}
\newcommand{\bel}{\begin{eqnarray*}\begin{aligned}}
\newcommand{\eel}{\end{aligned}\end{eqnarray*}}

\def\cB{\mathcal{B}} 
\def\theh{ih}\def\theh{t}

\def\PnL{PnL\xspace}

\def\thec{c}
\def\thenb{n}

\def\I{\mathds{1}}
\def\Id{\mathrm{diag}}

\newcommand{\indi}[1]{\I_{{#1}}}

\def\nabla{\partial}
\def\qqq{\;\;\;\;\;\;\;\;\;}
\def\R{\mathbb{R}} 

\def\si{\si }\def\si{i}

\def\tjp{\tjp }\def\tjp{j+1}

\def\sj{\sj }\def\sj{j}
\def\si{\si }\def\si{i}
     
 \def\Var{\mathbb{V}\mathrm{ar}}
\def\widetilde{}

\def\rhob{\bar{\rho}} 
\def\varrhob{\bar{\varrho}} 

\def\rmo{\rm o}\def\rmo{\omega}

\def\pointwise{deterministic}\def\pointwise{pointwise\xspace}
\def\bo {\textbf{0}}\def\bo{0}
\def\CF{\textrm{CF}}
\def\ot{0}\def\ot{(t)}
\def\PLE{PLE\xspace}
\def\LS{LS\xspace}
\def\smart{smart\xspace}
\def\Smart{Smart\xspace}
\begin{document} 
\title{CVA Sensitivities, Hedging and Risk
%\Smart Sensitivities and CVA Applications
\footnote{The python code of the paper is available on \url{https://github.com/botaoli/CVA-sensitivities-hedging-risk}.}
}
%\\Learning and Hedging the CVA\\Learning the CVA and Its Sensitivities \\Learning and Explaining the CVA 
\author{
S. Cr\'epey\thanks{Email: {\tt stephane.crepey@lpsm.paris}. LPSM/Universit\'e Paris Cité, 
France.  {\tt Corresponding author}. 
%A very preliminary (unpublished) version of this work has benefited from the support of the Research Initiative ``Mod\'elisation des march\'es actions, obligations
% et d\'eriv\'es''
%  financed by HSBC France under the aegis of the Europlace Institute of Finance.
  } , 
  B. Li\thanks{Email: {\tt botaoli@lpsm.paris}. LPSM/Université Paris Cité. The research of B. Li is funded by the Chair \textit{Capital Markets Tomorrow: Modeling and Computational Issues} under the aegis of the Institut Europlace de Finance,  a joint initiative of Laboratoire de Probabilit\'es, Statistique et Modélisation (LPSM) / Université Paris Cit\'e and  Cr\'edit Agricole CIB, with the support of Labex FCD (ANR-11-LABX-0019-01).}, 
H. Nguyen\thanks{Email: {\tt hdnguyen@lpsm.paris}. LPSM/Université Paris Cité. The research of H.D. Nguyen is funded by a CIFRE grant from Natixis.} , 
B. Saadeddine\thanks{Email: {\tt bouazza.saadeddine2@ca-cib.com}. Quantitative research GMD, Credit Agricole CIB, Paris.} 
}

{\let\thefootnote\relax\footnotetext{{{\it Acknowledgement:} We are grateful to Moez Mrad,
head of XVA, counterparty risk, collateral, and credit derivatives quantitative research at Crédit Agricole CIB, 
and to an anonymous referee, for inspiring exchanges.}}} 
% , and to Dorinel Bastide, quantitative analyst at BNP Paribas stress testing methodologies \& models, for precious
% regulatory insights.}}} 

\maketitle
%\vspace*{-0.3cm}
\begin{abstract} 
We present a unified framework for computing CVA sensitivities, hedging the CVA, and assessing CVA risk, using probabilistic machine learning meant as refined regression tools on simulated data, validatable by low-cost companion Monte Carlo procedures. Various
notions of sensitivities 
%and ways of computing them 
are introduced 
%conceptually 
and benchmarked numerically. We identify the sensitivities representing the best practical trade-offs in downstream tasks including CVA hedging and risk assessment.   

% As an application, we demonstrate that sensitivities-based approaches to CVA risk may be insufficient. A better view on CVA risk can be provided by the  
%  learned CVA fed by statistical views on CVA risk factors. 
% \Smart pricing has developed along two modes, interpolation of prices and regression of cash flows. 
% %The latter, rooted in the proven
% %% long and respectable
% % tradition of simulation/regression schemes for Bermudan options, increasingly emerges as a much more reasonable and promising avenue. 
% This work first contributes to the latter, in the companion area of \smart sensitivities. The effectiveness of our approach is illustrated on CVA computations. We also demonstrate that sensitivities-based approaches to CVA risk may be insufficient. A better view on CVA risk can be  provided by a %pricing engine trained to learn the CVA pathwise and
%  learned CVA fed by statistical %(or users') 
%  views on CVA risk factors. 

%Toolbox
\end{abstract}

\def\keywordname{{\bfseries Keywords:}}
\def\keywords#1{\par\addvspace\baselineskip\noindent\keywordname\enspace
\ignorespaces#1}

\section{Introduction}

This work illustrates the potential of probabilistic machine learning for pricing and Greeking applications, in the challenging context of  CVA computations. By probabilistic machine learning we mean machine learning as refined regression tools on simulated data.
Probabilistic machine learning for  CVA pricing was introduced in
\citet*{abbas2021hierarchical}.
Here we extend our approach to encompass CVA sensitivities and risk.
The fact that probabilistic machine learning is performed on simulated data, which can be augmented at will, does not mean that there are no related data issues.
As always with machine learning, the quality of the data is the first driver of the success of the approach. The variance of the training loss may be high and jeopardize the potential of a learning approach. This 
%difficulty 
was first encountered in the CVA granular defaults pricing setup of  \citet{abbas2021hierarchical} due to the scarcity of the default events compared with the  diffusive scale of the risk factors in the model. Switching from prices to sensitivities in this paper is another case of increased variance.
But with probabilistic machine learning  we can also develop
%tools to address these data issues, not only by simulating more data at will, but also by devising 
suitable variance reduction tools, namely oversimulation of defaults in  \citet{abbas2021hierarchical} and common random numbers in this work.  Another distinguishing feature of probabilistic machine learning,  which is key for regulated banking applications, is the possibility to assess the quality of a predictor by means of low-cost companion Monte Carlo procedures. 

\subsection{Outline of the Paper and Generalities}

Section
\ref{s:fs} introduces different variants of bump sensitivities, benchmarked numerically in a CVA setup in
Section \ref{s:thecva}.
Section \ref{s:cvalh} shows  how a conditional (e.g.\ future) CVA can be learned from simulated data (pricing model parameters and paths and financial derivative cash flows), both in a baseline calibrated setup and in an enriched setup also accounting for recalibration model shifts. 
Sections \ref{ss:riskdyn} and \ref{s:risk}
develop a framework for internal modeling of CVA and/or counterparty default risks, entailing various notions of CVA sensitivities. 
Section \ref{s:concl} concludes as for which kind of
sensitivity and numerical scheme provide the best practical trade-off for various downstream tasks including CVA hedging and risk assessment. 

A 
%row-
vector of partial derivatives with respect to $x$ is denoted by $\partial_x$ (or $\cdot'$ when $x$ is clear from the context).
All equations are written using
the risk-free asset as a num\'eraire and are stated under the probability measure
%, denoted by $\mathbb{R}$ for ``regulatory", 
which is the blend of physical and pricing measures advocated for XVA computations 
in \citeN[Remark 2.3]{albanese2021xva}, with related expectation operator denoted below by $\mathbb{E}$. %In particular, the PnLs of the different business desks (CVA desk included) of the bank are centered under this measure \citep*[Remark 2.4]{Crepey21}.
All cash flows are assumed square integrable.
Some standing notation is listed in Table \ref{t:nota}.
\begin{table}[b!]
\centering
\resizebox{\textwidth}{!}{
\begin{tabular}{|ll|ll|}
\hline
\begin{tabular}[c]{@{}l@{}}$\rho$\\ {\color[HTML]{FFFFFF} ee}\\ {\color[HTML]{FFFFFF} ee}\end{tabular} & \begin{tabular}[c]{@{}l@{}}model parameters, i.e. exogenous \\ model parameters $\epsilon$ and initial \\ conditions of model risk factors\end{tabular} & \begin{tabular}[c]{@{}l@{}}$Y$\\ {\color[HTML]{FFFFFF} ee}\\ {\color[HTML]{FFFFFF} ee}\end{tabular} & \begin{tabular}[c]{@{}l@{}}CVA diffusive risk factor process \\ with initial condition $y$ \\ (if constant) or $\iota$ (if randomized)\end{tabular} \\
$\varrho$ & randomization of $\rho$ & $X$ & default indicator processes of clients \\
\begin{tabular}[c]{@{}l@{}}$\omega$\\ {\color[HTML]{FFFFFF} ee}\end{tabular} & \begin{tabular}[c]{@{}l@{}}stochastic drivers \\ (e.g.\ Brownian paths)\end{tabular} & \begin{tabular}[c]{@{}l@{}}$Z$\\ {\color[HTML]{FFFFFF} ee}\end{tabular} & \begin{tabular}[c]{@{}l@{}}market price process, with \\ initial condition $z_0$\end{tabular} \\
$\rho_0$ & calibrated value of $\rho$ & $\xi(\rho; \omega)$ & product payoff \\
$\bar{\cdot}$ & $2\rho_0-\cdot$ & $\Pi_0$ & $\E \xi ( \rho ) $ \\
$\cdot^\theta$ & neural net function with parameters $\theta$ & $\varsigma(\rho; \omega)$ & $ \xi ( \rho ;\omega )-\xi ( \rhob;\omega )$ \\
$q$ & number of market instruments & $\Sigma_0( \rho )$ & $\E \varsigma ( \rho ) $ \\
\begin{tabular}[c]{@{}l@{}}$m$\\ {\color[HTML]{FFFFFF} ee}\end{tabular} & \begin{tabular}[c]{@{}l@{}}number of Monte Carlo paths \\ of the pricing model\end{tabular} & \begin{tabular}[c]{@{}l@{}}$\Delta$\\ {\color[HTML]{FFFFFF} ee}\end{tabular} & \begin{tabular}[c]{@{}l@{}}CVA hedging ratio or linear \\ coefficients of a CVA proxy\end{tabular} \\
\begin{tabular}[c]{@{}l@{}}$p$\\ {\color[HTML]{FFFFFF} ee}\end{tabular} & \begin{tabular}[c]{@{}l@{}}number of model parameters, \\ i.e.\ dimension of $\rho$\end{tabular} & \begin{tabular}[c]{@{}l@{}}$\Gamma$\\ {\color[HTML]{FFFFFF} ee}\end{tabular} & \begin{tabular}[c]{@{}l@{}}diagonal quadratic coefficients \\ of a CVA proxy\end{tabular} \\ \hline
\end{tabular}}
\caption{Standing notation.}\label{t:nota}\vspace*{-0.5cm}
\end{table}

All neural network trainings are done using the PyTorch module and the Adam optimizer.
%implemented in PyTorch along with CUDA kernels on GPU for the generation of the CVA originating cash flows at the forward simulation stage,
Linear regressions are implemented using a truncated singular value decomposition (SVD) approach. 
%(see~e.g.~\citeN[Theorems 2.5.2 and 5.5.1]{golub2013matrix}). 
Unless explicitly stated, we always include a ridge (i.e.\ Tikhonov) regularization term in the loss function to stabilize trainings and regressions.
Our computations are run on a server with an Intel(R) Xeon(R) Gold 5217 CPU and a Nvidia Tesla V100 GPU.

\FloatBarrier
%\part{Bump Sensitivities}

\section{Fast Bump Sensitivities}% \b{\Smart Regression Sensitivities}
\label{s:fs}
In this section we consider a time-0 option price
$\Pi_0( \rho   )  =\E  \xi ( \rho   ),$ 
% \b{should it be better with $\Pi_0 ( \rho   )= \E  \big[\xi ( \varrho , \omega  ) \mid \varrho = \rho\big]$  ?},
where the 
%$\F_T$ measurable
payoff 
$  \xi (\rho    ) \equiv \xi (\rho ;  \omega   )$ depends
on  constant model parameters $\rho$ and (implicitly in the shorthand notation  $  \xi ( \rho   )$) on the randomness $\omega$ of the stochastic drivers of the model risk factors with respect to which the expectation is taken above.
The model parameters $\rho$ encompass the
initial values of the risk factors of the pricing model, as well as all the exogenous (constant, in principle) model parameters, e.g.\ the value of the volatility in a Black-Scholes model.
For each constant $ \rho $,
the price $\Pi_0( \rho   )$ can be estimated by Monte Carlo. Our problem in this part is the estimation of the corresponding sensitivities $ \partial_\rho \Pi_0 (\rho_0) $,
at a baseline (in practice, calibrated) value $ \rho=\rho_0  $
of the model parameters.   
Such sensitivities lie at the core of any related hedging scheme for the option. They are also key in many regulatory capital formulas.
%\b{(for the CVA capital charge in particular, see Section \ref{s:risk})}.
%CITE DORI}.
%the regulatory  standard approach for the CVA capital charge \citep[Articles 383i--j pages 174--178]{EU21}.   %\b{Hereafter, we use the acronym ``sensis'' for the word ``sensitivities''.}

Monte Carlo estimation of sensitivities in finance comes along three main streams \citep[Section 6.6]{cre}: (i) differentiation of the density of the underlying process via integration by parts or more general Malliavin calculus techniques, assuming some regularity of this process; (ii) cash flows differentiation, assuming their differentiability, in chain rule with the stochastic flow of the underlying process;  (iii) Monte Carlo finite differences, biased but generic, which are the Monte Carlo version of the industry standard bump sensitivities. 
But (i)
%Malliavin calculus reformulations of payoff sensitivities 
suffers from intrinsic variance issues. 
In contemporary technology, 
%payoff sensitivities 
(ii)
appeals to adjoint algorithmic differentiation (AAD). A randomized version of this approach
%(ii) 
is provided by Sections 5.3--5.5 of \citeN{saadeddine2022learning},
%As illustrated numerically in its Section 5.5, 
%The corresponding approach 
targeted to model calibration, which requires sensitivities of vanilla options as a function of their model parameters. However, the embedded AAD layer 
can quickly represent important implementation and memory costs on complex pricing problems at the portfolio level such as CVA computations: see \citet*{capriotti2017aad}. 
% AAD through a training routine to differentiate an (approximated) inner conditional expectation
% As ``compound AAD'' is unpractical, 
Such an AAD Greeking approach becomes nearly unfeasible in the case of pricing problems embedding numerical optimization subroutines, e.g.\ the training of the conditional risk measures
%auxiliary XVA layers
embedded in the refined CVA and higher-order XVA metrics of \citet{albanese2021xva} (with Picard iterations) or \citet*{AbbasturkiCrepeyLiSaadeddine23} (explicit scheme without Picard iterations).

 % In what follows we propose a learning twist on (iii), whereby a linear or more general neural network model of the finite differences $\Sigma_0(\rho)$ of the option price 
 % %(as opposed to the price itself) 
 % with respect to $\rho$ is learned and yields an estimate of the sensitivities $\partial_{\rho}\Pi_0(\rho_0)$ as a by-product. When used
 % in conjunction with appropriate distributions for the embedded randomization $\varrho$ of $\rho$, this approach can diminish significantly the computational burden for the first-order bump sensitivities, especially when there are many of them involved and pricing can only be achieved by intensive Monte Carlo simulations, as in CVA computations. The ensuing model sensitivities can then be turned into market sensitivities, i.e.\ sensitivities to the prices of suitable calibration/hedging instruments, by classical Jacobian transformations.

\subsection{Common Random Numbers}\label{s:fast_ss}
%Key Idea

Under the approach (iii),
first-order \pointwise bump sensitivities are computed by relaunching the Monte Carlo pricing engine
with common random numbers $\omega$ 
%(see Section \ref{s:fast_ss})
for values bumped by $\pm 1\%$ (typically and in relative terms) of
each risk factor and/or model parameter of interest, then taking the accordingly normalized difference between the 
corresponding $\Pi_0(\rho)$ and
$\Pi_0(\rhob)$, where
$\bar{\cdot}$ means symmetrization with respect to $\rho_0$,
so
\beql{e:sym}\frac{  \rho+\rhob}{2} =\rho_0\sp \mbox{i.e. } \rhob=2\rho_0-\rho .\eeql
This approach requires two Monte Carlo simulation runs per sensitivity, making it a robust but heavy procedure, which we try to accelerate by various means in what follows.

A predictor $\Pi_0^{\theta}(\rho)\approx\Pi_0(\rho) $ for the pricing function 
around $\rho= \rho_0$
within a suitable space of neural nets parameterized by $\theta$
%(or linear functions as a limiting case) for $\rho \approx \rho_0$ 
readily leads to an AAD 
%(or even simpler extraction in the linear case) 
estimate $\partial_\rho \Pi_0 (\rho_0) \approx  \partial_\rho \Pi^{\theta}_0(\rho_0)$ for the corresponding sensitivities. 
However, even if ridge regularization may help in this regard, such an estimate, deemed naive AAD hereafter, may be bad as differentiation is not a continuous operator in the supremum norm (in other terms, functions may be arbitrarily close in  sup norm but their derivatives may be far from each other). 
Specifically, let $\cB  $ denote the space of the Borel measurable functions of $\rho$.
The pricing function \beql{e:lepi}\Pi_0( \rho)= \E \xi( \rho)= \E\big(\xi   ( \varrho  )  \,\big|\,  \varrho= \rho \big) \eeql 
(for $\varrho$ randomizing $\rho$ around $\rho_0$) can be %efficiently 
learned from simulated pairs $(\varrho,\xi(\varrho;\omega))$ based on the representation 
%(assuming square integrable cash flows $\xi$)
\beql{e:learnpi} \Pi_0 (\cdot) = \argmin_{\Phi\in\cB } \mathbb{E}\Big[\big(\xi    ( \varrho  )-\Phi(\varrho)\big)^2\Big].\eeql
To learn the function $\Pi_0(\cdot)$ around $\rho_0$, 
we can
replace, in the optimization problem 
\eqref{e:learnpi}, $\cB
  %( {\mathbb{R}^{q}} )
  $ by 
 a suitable space of neural nets and $\mathbb{E}$ by a simulated sample mean $\widehat{\mathbb{E}}$, with each ``vertical'' (time-0) draw of 
 %a suitable randomization 
 $\varrho$ 
 %of $\rho$ around $\rho_0$
 %(see line 1 of Algorithm \ref{a:fast_sensi})
 followed by an ``horizontal'' (across future times)
 %(over $(0,T]$) 
 draw of $\omega$ that is implicit in $\xi(\varrho)\equiv \xi(\varrho;\omega)$. 
The ensuing minimization problem for the weights $\theta$ of the neural net $\rho\mapsto 
\Pi^\theta_0(\rho)\approx \Pi_0(\rho)$ is then performed  numerically by Adam mini-batch  gradient descent.
The corresponding sensitivities $\partial_\rho \Pi^\theta_0 (\rho_0)$ are retrievable by AAD at negligible  additional cost.
However, as emphasized above, the ensuing \textbf{naive AAD sensitivities}
$\partial_\rho \Pi^\theta_0  $ may be a poor estimate of  $\partial_\rho \Pi_0  $. 

In this learning setup,
the corresponding instability reflects a variance issue.
% it is well known that the derivative of an approximation does not necessarily approximate well 
%Instead,
%Specifically, 
In order to cope with the increased variance due to the switch from prices to sensitivities, a useful trick is to introduce
$\varsigma(\rho;\omega):= \xi ( \rho ;\omega )-\xi ( \rhob;\omega  )$ (cf.\ \eqref{e:sym}).
We can then learn the sensitivity (in the sense here of finite differences) function 
$\rho\mapsto\Sigma_0( \rho  ) := \E \varsigma ( \rho  )  $, 
which satisfies by linearity and chain rule (as $\rhob=2\rho_0-\rho$)
\beql{e:half}
 \Sigma'_0 ( \rho  ) =  \Pi'_0( \rho   )
+ \Pi'_0( \rhob  )
%+\partial_{\rho}\Pi_0( \rhob  )
\mbox{, in particular } \partial_{\rho}\Sigma_0( \rho_0  ) =2 \partial_{\rho}\Pi_0( \rho_0  ).\eeql
For learning the function $\Sigma_0(\cdot)$ locally around $\rho_0$, with
$ \varrho$ randomizing $\rho$ as above, 
we rely on the representation
%\beql{e:thetaprel}
$\Sigma_0( \rho  ) = \E\big(\varsigma    ( \varrho  )  \,\big|\,  \varrho= \rho \big) ,$
%\eeql 
i.e.
\beql{e:theta} \Sigma_0 (\cdot) = \argmin_{\Phi\in\cB } \mathbb{E}\Big[\big(\varsigma    ( \varrho  )-\Phi(\varrho)\big)^2\Big].\eeql 
Then we
replace, in the optimization problem 
\eqref{e:theta}, $\cB
  %( {\mathbb{R}^{q}} )
  $ by 
 a linear hypothesis space $\Sigma^\theta_0(\rho)=\theta^\top (\rho-\rho_0)$  (noting that $\Sigma_0(\rho_0)=0$) and $\mathbb{E}$ by a simulated sample mean $\widehat{\mathbb{E}}$, with again each vertical draw of $\varrho$ followed by one horizontal
 %(over $(0,T]$) 
 draw of $\omega$ that is implicit in $\varsigma(\varrho)$. 
This results in a linear least-squares problem for the weights $\theta$, solved by SVD.
% The corresponding linear regression is readily solved by 
% singular value decomposition (SVD, see~e.g.~\citeN[Theorems 2.5.2 and 5.5.1]{golub2013matrix}). 
The estimated weights $\theta/2$
%$\hat{\varphi}$ 
are our 
%\smart 
\textbf{linear bump sensitivities} estimate for $\frac{1}{2} \partial_\rho \Sigma_0 (\rho_0)=  \partial_{\rho}\Pi_0(\rho_0 )$. 
These sensitivity estimates 
are the slope coefficients of a multilinear regression, for which
confidence intervals CI scaling in $1/\sqrt{| \mbox{sample size} |}$ are available %\b{more recent book ref?}%\citep[Theorem 3]{White1980}.
%\citep[section 12.4.1]{DingBook} 
\citep[Section 2.12.11]{matloff2017statistical}.
%\r{Not what we really use, but should be fine as a first introduction of CI}.
The use of each drawn set of model parameters $\varrho$ twice, also via $\varrhob$  with a common $\omega$ in 
$\varsigma(\varrho;\omega)=\xi(\varrho;\omega) -\xi ( \varrhob ;\omega )$,  
is a common random numbers variance reduction technique as in (iii) above. For well-chosen distributions of the randomization $\varrho$ of $\rho$,
this approach results in much more accurate sensitivities than the naive AAD approach.
% , at the cost of a single run of the  Monte Carlo pricing engine plus a linear regression, versus one Monte Carlo run of the pricing engine \textit{per sensitivity} in the case of \pointwise bump sensitivities. 
For simple parametric distributions of $\varrho$, the covariance matrix that appears in the regression for the first-order sensitivities
is known and invertible in closed form, which reduces the linear regression (implemented without ridge regularization in this analytical case) to a standard Monte Carlo and
a more robust CI.
Nonlinear hypothesis spaces of neural networks trained the way described after \eqref{e:learnpi} 
(just replacing $\xi(\varrho)$ and $\partial_\rho \Pi^\theta_0 (\rho_0)$
there by $\varsigma(\varrho)$ and $\partial_\rho \Sigma^\theta_0 (\rho_0)$ here) can also be used instead of the above linear model for $\Sigma_0(\rho)$. The \textbf{AAD bump sensitivities} are then obtained as the halved AAD sensitivities of the trained (no longer linear)  network $\rho\mapsto \partial_\rho \Sigma^\theta_0
% (\rho)
$ at $\rho_0$ (but we lose the confidence interval CI in this case). Similar ideas can be applied to higher order sensitivities, using e.g. %\beql{e:findd} 
$ \xi ( \rho  ;\omega)-2 \xi ( \rho_0;\omega  ) +\xi ( \rhob ;\omega )$
%\eeql
instead of $\varsigma(\rho;\omega) = \xi ( \rho ;\omega )-\xi ( \rhob;\omega  )$
to capture diagonal gammas. However, higher order means even more variance. Moreover, the Jacobian trick of Section \ref{ss:jac}  to convert model into market sensitivities is only workable for first-order sensitivities, hence our focus on the latter in this work.
 
An important ingredient in successful randomized (linear or AAD) bump sensitivities is the choice of appropriate distributions for $\varrho$. Since different parameters (components of $\rho$) may have very different magnitudes in values and price impact  (see e.g.\ Figure \ref{fig:bs} page \pageref{fig:bs} and Table \ref{t:sensis} page \pageref{t:sensis}), a linear regression may not be able to identify the individual effect of each  parameter  when  all are  bumped simultaneously. 
% In particular, shocking all parameters at once produces unstable estimates due to inversion of the large nonsparse 
% %même si diagonale asymptotiquement comme estimant la mat de cov de facteurs indépendants
% %%\b{ill-conditioned}
% matrices $(\delta \rho)^\top \delta \rho $, where by $\delta \rho$ we mean the $p\times m$ matrix with columns $\rho_j-\rho_0, j\in 1\,..\,m$. 
% %depending on both $m$ \b{and $p:=|\rho|$}. %\g{$(X^\top X)^{-1}$ where $X$ is $m\times p$} 
To address this issue, we divide the parameters into groups. The $m$ simulated paths of the pricing model are then partitioned into blocks of paths such that only one group of parameters is bumped in each subset. 
%?bib blocs Pesquet chouzenoux  or
In addition to this, one can use different distributions
%noise scales $\sigma$ in Algorithm \ref{a:fast_sensi} 
for each group or even for each parameter. Notably, we have observed numerically
that slightly 
noiser distributions
%larger noise scales 
yield better sensitivities estimates 
for volatility-related parameters.
\hspace{-0.65cm}\begin{figure}[t!]
\begin{minipage}[b!]{\textwidth}
\begin{algorithm}[H] 
\small
\LinesNumbered
\SetAlgoLined
%\SetKwInOut{AlgName}{name}
\SetKwInOut{Input}{input}\SetKwInOut{Output}{output}
%\AlgName{BayesRegSensi}
\Input{A (calibrated) baseline $\rho_0$ for the initial conditions of all risk factors and for the exogenous parameters of the pricing model,
%a stochastic model parameter diffusion, 
 a number $m$ of pricing model paths.
}
\Output{Estimated sensitivities $\theta/2$ (linear case) or $ \frac{1}{2}\partial_\rho \Sigma^\theta_0 (\rho_0)$ (more generally) $\approx\partial_\rho \Pi_0 (\rho_0)$.} 
Draw $m$ i.i.d. bumped model parameters $
%\mathrm{P} = 
\rho_j$  from some distribution randomizing $\rho$ around $\rho_0$, e.g.\ $\mathcal{N}\Big(\rho_0, \Id \left(\sigma^2\rho_0 \odot \rho_0 \right) \Big)$\footnote{$\odot$ is the Hadamard (i.e.\ componentwise) product between vectors and $\Id({\rm vec})$ is a diagonal matrix with diagonal ``vec''.} with $\sigma = 1\%, 3\%, 5\%$\footnote{The notation $ \mathcal{N}\Big(\rho_0, \Id \left(\sigma^2\rho_0 \odot \rho_0 \right) \Big)$ which is used for simplicity in this pseudo-code ignores the practically important block simulation trick mentioned in the end of Section \ref{s:fast_ss}.} or, in the case of smart bump sensitivities, $\rho_j = \rho_0$ bumped by  $1\%$ on its $\left\lfloor  \frac{jp}{m} \right\rfloor$-th component  

Draw $m$ i.i.d. stochastic drivers $\omega$ parameters $
%\mathrm{O}  = 
\{\rmo _1,\dots, \rmo _{m}\}$

 \For%(\tcp*[f]{loop over model parameter set})
 {$(\rho,\rmo ) \in  (\rho_1,\rmo _1)\,..\, (\rho_1,\rmo _{m})
 %(\mathrm{P}, \mathrm{O}) 
 $}{
    Compute the payoffs $\xi( \rho; \rmo  )$ and $\xi( \rhob; \rmo  ) $, where $\frac{\rho + \rhob}{2} = \rho_0$\\
    Compute $\varsigma(\rho;\rmo ) = \xi ( \rho; \rmo   )-\xi ( \rhob ; \rmo  )$
 }
\uIf{AAD bump sensitivities}{
Train a neural network $\Sigma_0^\theta(\rho)$ to regress $(\varsigma(\rho_1),\dots, \varsigma(\rho_m))$ against $(\rho_1, ,\dots, \rho_m)$\\
%Predict $\varsigma(\rho_0)$, 
Retrieve $\partial_\rho\Sigma_0^\theta(\rho)$ by AAD and divide the obtained gradient by two for obtaining the AAD bump sensitivities.}
\uElseIf{Linear bump sensitivities}{
Regress linearly (without intercept) $(\varsigma(\rho_1),\dots, \varsigma(\rho_m))$ against $(\rho_1 -\rho_0 ,  \dots, \rho_m-\rho_0)$ by SVD  \\
Divide the regression coefficients $\theta$ by two for obtaining the linear bump sensitivities% estimates 
}
\ElseIf{\Smart bump sensitivities}{
Compute each sensitivity by dividing the average of $\varsigma$ over each block of size $m/p$ by two times the corresponding  bump size
}
\caption{Fast bump sensitivities.}
\label{a:fast_sensi}
\end{algorithm}
\end{minipage}
\end{figure}

In addition to the above, we will also compute
\textbf{benchmark bump sensitivities} as per (iii) in the above, obtained on the basis of $p$ Monte Carlo repricings with $m$ common random numbers $\omega$ and relative variations of $\pm 1\%$ of one model parameter in each Monte Carlo run, as well \textbf{\smart} %(pointwise) 
\textbf{bump sensitivities}, similar but only using $m/p$ paths each, 
where $p$ is the number of model parameters (dimension of $\rho$). Hence the time of computing all the \smart bump sensitivities is of the same order of magnitude as the one of
retrieving the linear bump sensitivities, which is also roughly the time of pricing $\Pi_0(\rho_0)$ by Monte Carlo with $m$ paths. More precisely,
each \smart bump sensitivity uses $m / p$ paths of a Monte Carlo simulation run with $m$ paths as a whole. This is significantly more efficient 
%to simulate the \CVA payoffs 
than doing $p$ Monte Carlo runs of size $m / p$ each, especially in the GPU simulation environment of our CVA computations later below. Note that such smart bump sensitivities are also a special case of linear bump sensitivities with block simulation trick, for blocks of size $m/p$ and deterministic bumps of relative size 1\% on one and only one model parameter in each block, the linear regression degenerating in this case to a local sample mean over the $m/p$ paths of each block: see Algorithm \ref{a:fast_sensi}, which summarizes the above procedures for various fast bump sensitivities.

\FloatBarrier
\subsection{Basket Black-Scholes Example}\label{s:bs} 

Let us consider a European call on the geometric average of  $d$ Black-Scholes assets, aiming for the corresponding deltas, vegas, and (diagonal) gammas, which are known analytically in this lognormal setup. Regarding the above block simulation trick, the time-0 values of the assets are bumped in half of the paths, while the volatilities are bumped in the other half. We study two cases, with $d=3$ and $10$ and, respectively $m=10^5$ and $m = 5\times 10^5$  simulated Black-Scholes paths. In the $d=3$ case, we implement all the fast bump approaches of Algorithm \ref{a:fast_sensi} and the naive AAD approach described after \eqref{e:learnpi}. The learning of the gammas is also reported in this low-dimensional experiment. In the $d=10$ case, we skip the naive AAD approach as well as all the learned gamma results because of their poor performance. The hypothesis space used in all the AAD approaches is a vanilla multi-layer perceptron with two hidden layers and softplus activation functions. 
% Alternative learning models including richer neural networks or boosted trees have been tested without tangible benefit in all our use-cases in this work, hence we stay with MLPs, just named neural networks hereafter. 

\begin{figure}[!t]
% \resizebox{\textwidth}{!}
% %{
%\vspace*{-1cm}
%\hspace*{-1cm}
\begin{subfigure}{\linewidth}
 \hspace*{0.4cm}
 \includegraphics[width=0.34\linewidth, height = 170pt]{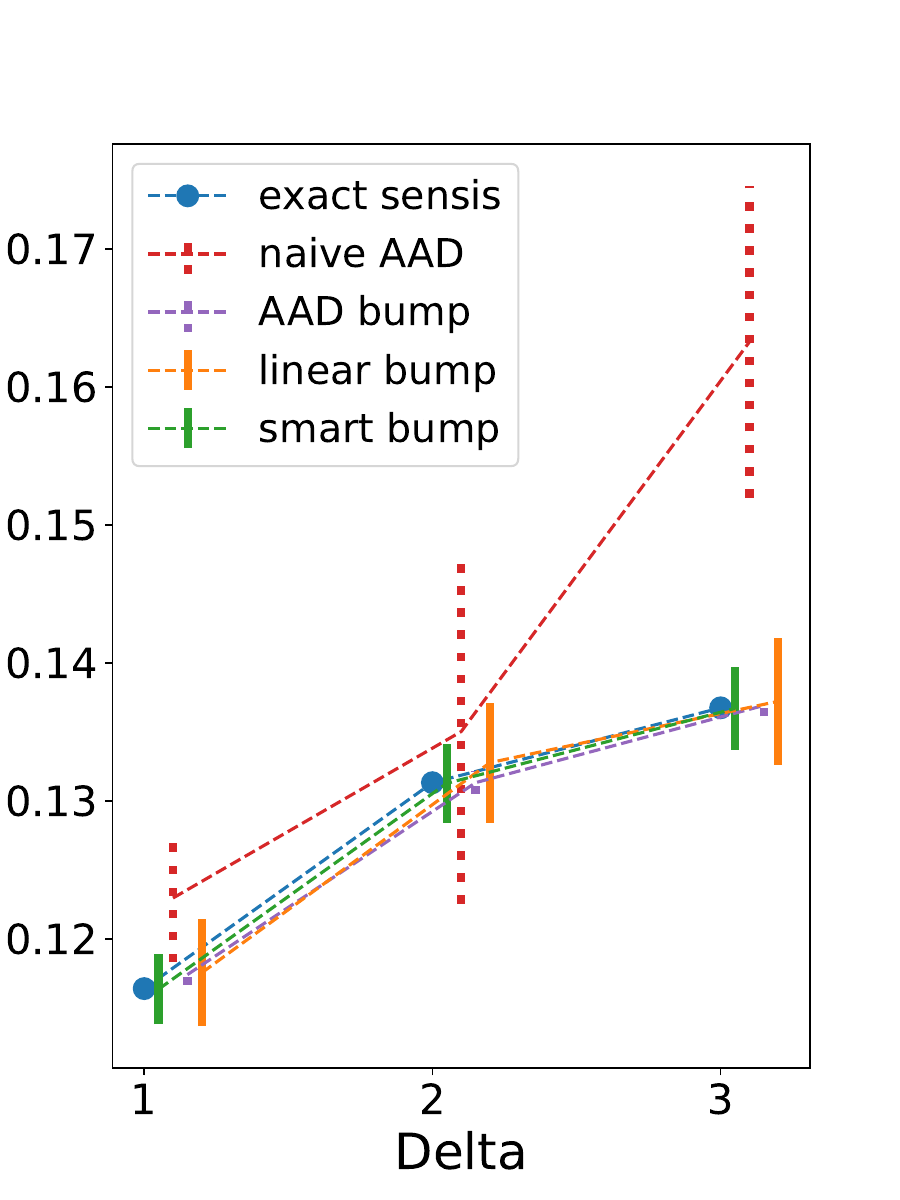}\hspace*{-0.3cm}
 \includegraphics[width=0.34\linewidth, height = 170pt]{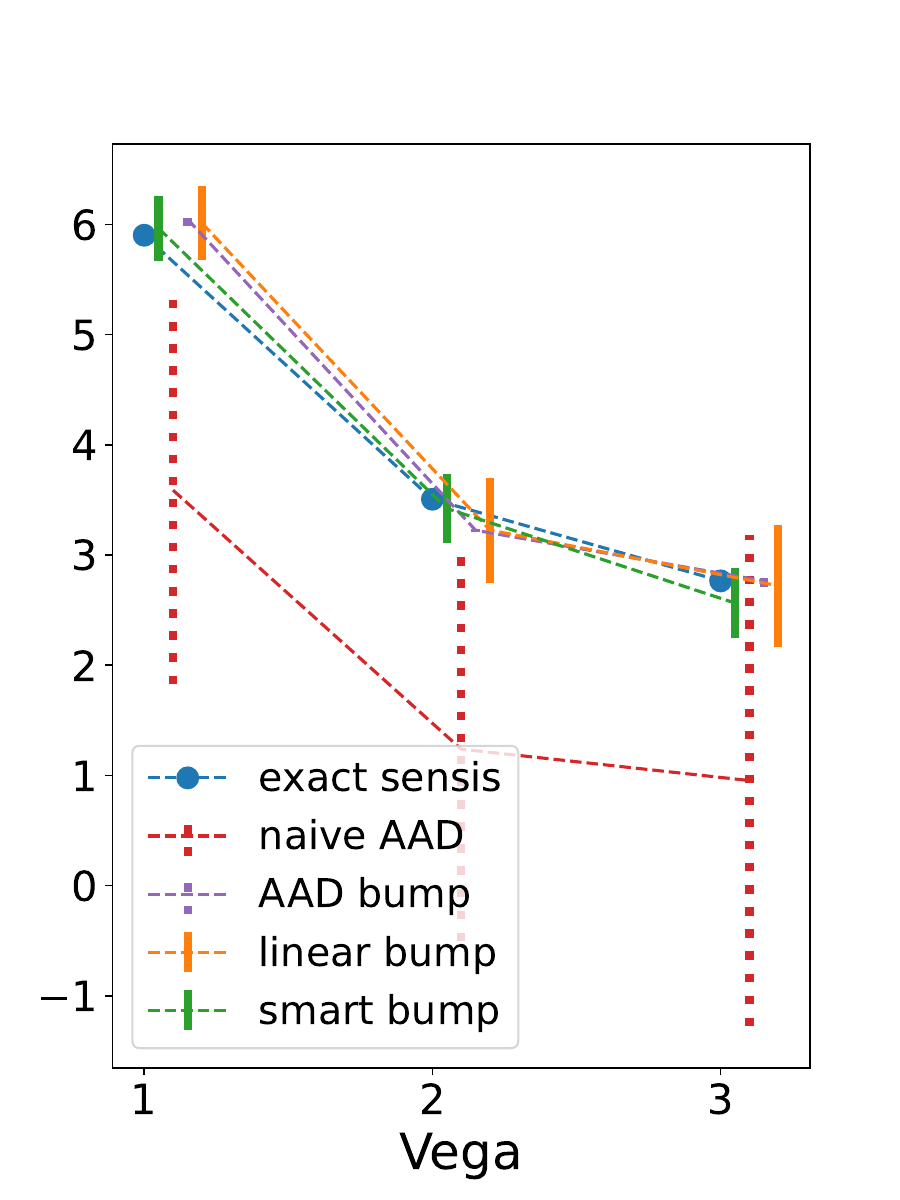}\hspace*{-0.3cm}
 \includegraphics[width=0.34\linewidth, height = 170pt]{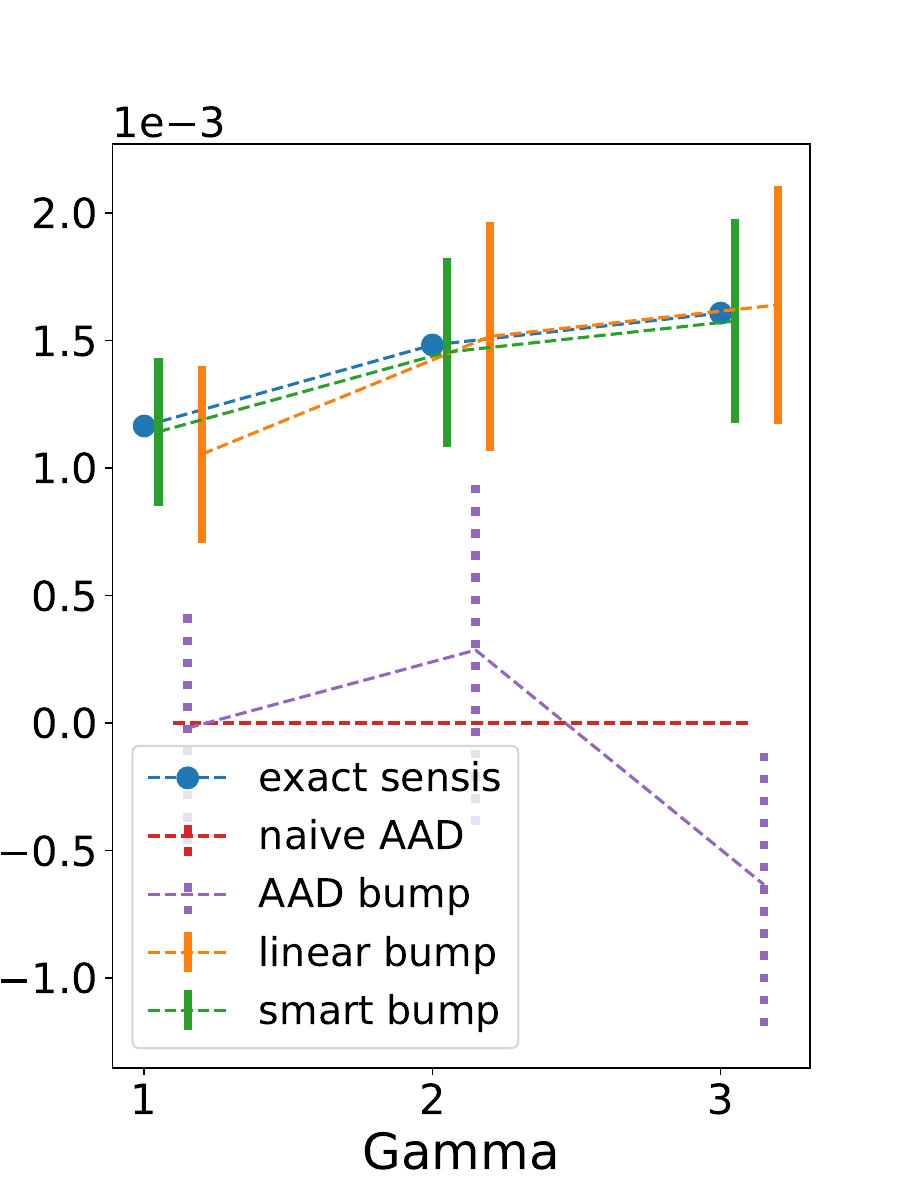}
 \caption{$d = 3$ and $m = 10^5$ paths.}
\end{subfigure}
\begin{subfigure}{\linewidth}
%\vspace*{-1.3cm}
 \includegraphics[width=0.55\linewidth, height = 170pt]{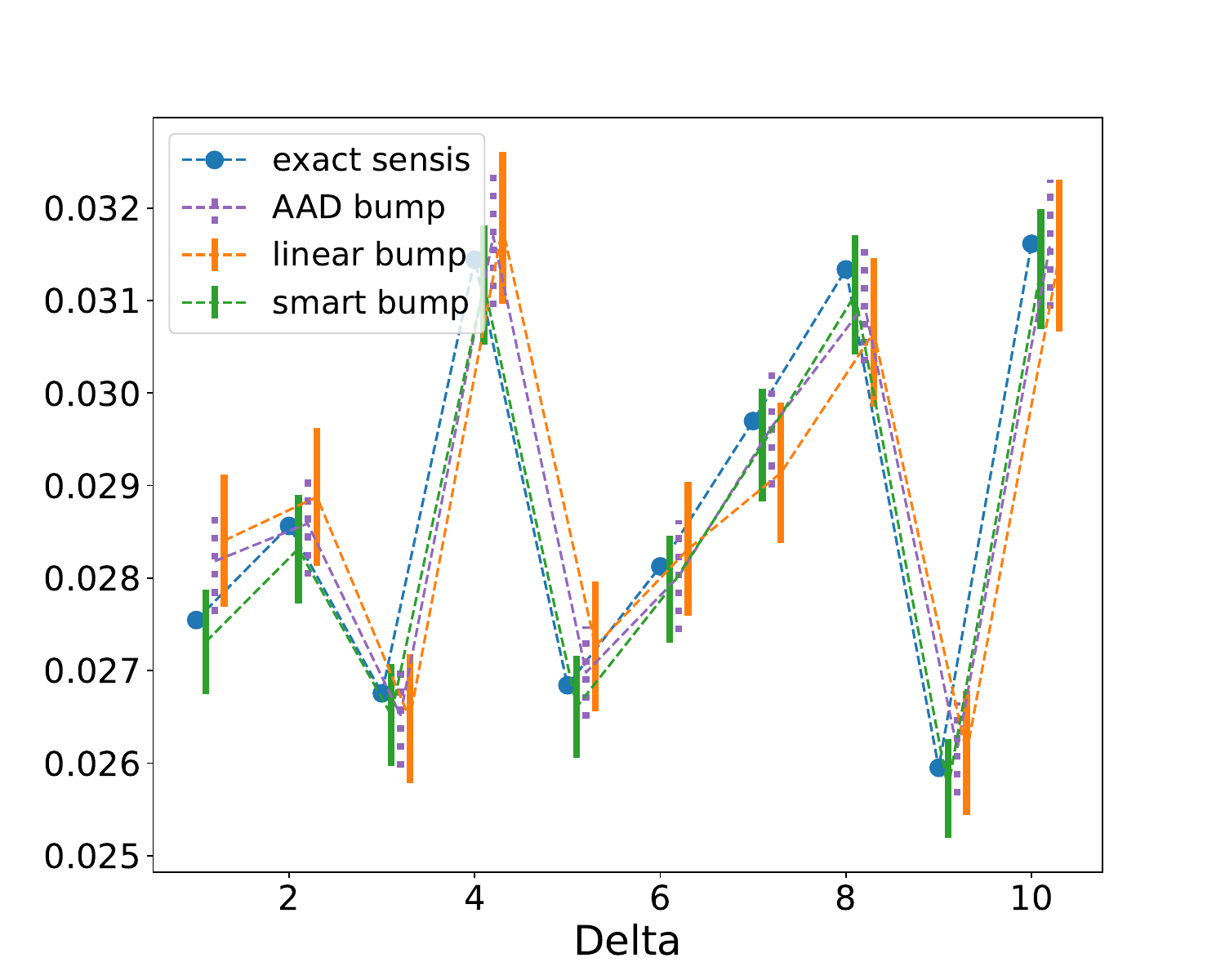}\hspace*{-0.8cm}
 \includegraphics[width=0.55\linewidth, height = 170pt]{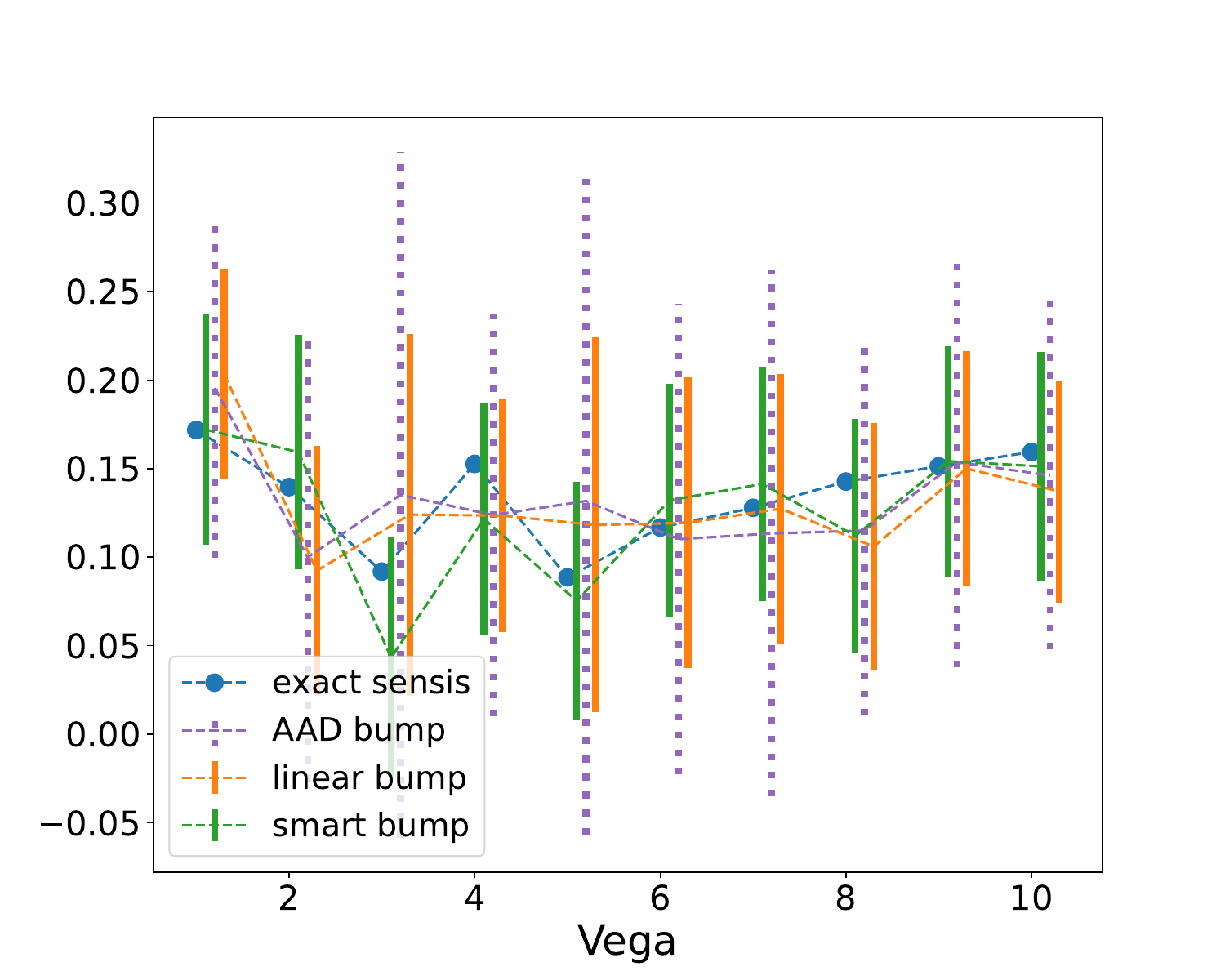}
 \caption{$d = 10$ and $m = 5  \times10^5$ paths.}
\end{subfigure}
\caption{%\b{change name and order: exact sensis (blue), naive AAD (red), AAD bump (violet), linear bump (orange), smart bump (green)} 
Multi-asset geometric average call sensitivities on 3 (a) and 10 (b) independent Black-Scholes assets. The error bars represent confidence intervals CI or CI$^\flat$, in dotted lines regarding the latter, which only reflect a training noise. }
\label{fig:bs}
\end{figure}

The  error bars in Figure \ref{fig:bs} page \pageref{fig:bs} represent our $95\%$ confidence intervals (CI) for the linear bump sensitivities. Regarding the AAD bump approach, we train the neural network %approximating $\varsigma$ in \eqref{e:theta}, see also line 11 in Algorithm \ref{a:fast_sensi}, 
100 times with different initializations to also get a 95\% confidence interval, in a meaning weaker than CI though: as the data is not resampled in each run, the AAD confidence intervals (CI$^{\flat}$) can only account for the randomness in training, not for the one of the simulated data. 
The upper plots of Figure \ref{fig:bs} show the inaccuracy of the naive AAD sensitivities.  
% We also observe that the exact sensitivities are aligned with the linear bump sensitivities and always fall within their confidence intervals CI, which is not the case for the AAD bump confidence intervals CI$^\flat$.  
The fast bump approaches consistently estimate deltas, but vegas and gammas appear to be more challenging. This is due to greater variance in the case of gammas, whereas in the case of vegas, the collusion between the noise of the random volatility coefficient and the one of the Brownian drivers $\omega$ makes the learning task more difficult.  
As should be, the CIs of the linear and smart bump sensitivities contain nearly all the exact sensitivities. This is also mostly the case for the CI$^\flat$s of the AAD bump sensitivities, but in their case this comes without theoretical guarantee.
The benchmark bump sensitivities are exact with 2 significant digits (at least for deltas and vegas) in this Black-Scholes setup. They would be visually indistinguishable of the exact sensitivities if we added them on the graphs of Figure \ref{fig:bs}. The time of computing the \smart bump sensitivities is of the same order of magnitude as the one of
%\b{pricing $\CVA_0(\rho_0)$ by Monte Carlo with $m$ paths or of} 
retrieving the linear bump sensitivities, 
with also similar accuracy as demonstrated in Figure \ref{fig:speedup}(a) page \pageref{fig:speedup}, where both ratios
between running times and errors of the linear and \smart bump sensitivities with respect to the benchmark bump sensitivities
stay close to each other. The complexities, describing how long it would take for each algorithm to achieve a relative error in the Black-Scholes case (or standard error in the CVA case) of 1\%, are not significantly different for the linear, \smart and benchmark bump sensitivities, for most of $p$; otherwise the \smart ones beat  by a small margin the linear ones, themselves a bit {worse} than the benchmark ones. %, \smart bump sensitivities 
%\b{(optimally implemented the way detailed in the end of Section \ref{s:fast_ss})} 
%beating  by a small margin linear bump sensitivities, \b{themselves a bit better\r{worse} than the benchmark sensitivities}. 
%The confidence interval of error ratios and complexities are determined 
%approximately 
%by the principle of uncertainty propagation of \citet[Chapter 4 page 22]{lee2005analyzing}, based on the result of $64$ independent runs. 
The grey dashed lines in the right panels indicate the $p^2$ scaling of complexities expected for the benchmark  bump sensitivities, which involve $2p$ simulation runs in dimension  $\Theta(p)$ (the number of risk factors in the pricing model).

In general, we observed that in high dimension the linear and \smart bump sensitivities tend to be more stable and reliable than the AAD bump sensitivities. Hence we forget AAD bump  
sensitivities hereafter.
%\FloatBarrier
\subsection{From Model to 
%Mark-to-
Market Sensitivities\label{ss:jac}}
{\def\star{}
\def\phi{\rho}
\def\varphi{\rho}
\def\vartheta{\psi}\def\vartheta{\rho}
\def\theta{\phi}\def\theta{\rho}
\def\pi{p}\def\pi{z}
\def\CVA{\Pi}
The sensitivities  $\partial_\rho \Pi_0 (\rho_0)$ are sensitivities to model  parameters. Practical hedging schemes require sensitivities to calibrated prices of hedging instruments. Hence, for hedging purposes, our sensitivities must be mapped
to hedging ratios in market instruments. This can be done via  
the implicit function theorem, the way explained in \citet{henrard2011adjoint}, \citet{savine2018model}, and \citet{AntonovIssakovMcClelland2017}.
 %We refer to  the conclusion of \citep{savine2018model} regarding efficient Jacobian transformations from the ones to the others, as well as on the inefficiency of direct derivation of sensitivities to market instruments.}
In a nutshell, assume that, given market prices $z\in \R^q$ of suitable calibration (hedging) assets, the corresponding pricing model parameters $\varphi(z)$ are obtained by a calibration procedure of the form
\begin{equation}\label{calibration_problem}
\theta^{\star}(z) \in \underset{\vartheta \in \mathbb{R}^p}{\mathrm{argmin}} \, cal\mbox{-}err\left( \pi,\vartheta\right), 
\end{equation}
where $cal\mbox{-}err\left(\pi,\vartheta\right)$ quantifies the 
%calibration error 
mean square error between the market prices $\pi$
of the calibration instruments and their prices in the pricing model with parameters $\vartheta$.  
Note that, in practice, not all the pricing parameters are obtained via a minimization as in \eqref{calibration_problem}: some of them are bootstrapped or even directly observed on the market, see \citet{savine2018model} for a more detailed presentation and Section \ref{s:toymo} for an example.
% In \eqref{calibration_problem} we restrict attention to the subset $\varphi\subseteq \rho$ of the parameters that are used for fitting the market prices $p$, the residual parameters being considered as fixed exogenously (see e.g.\ ).

Assume $cal\mbox{-}err$ of class $\mathcal{C}^2$ in $\vartheta$.
Let $\theta(\pi_0)=\theta_0$ be a solution of \eqref{calibration_problem} associated with a particular $\pi=\pi_0$, hence $\nabla_\varphi cal\mbox{-}err(\pi_0, \phi_0)=0$. 
Denote by $ \partial^2_{\vartheta ,\vartheta } cal\mbox{-}err   $ the Hessian matrix of $cal\mbox{-}err$ with respect to $\vartheta$ and assume that $ \partial^2_{\vartheta ,\vartheta } cal\mbox{-}err(z_0, \rho_0)   $ is invertible. Then, by the implicit function theorem applied to the function $\nabla_\varphi cal\mbox{-}err$ of $(z,\rho)$,  there exists an open neighborhood $\mathcal{O} \in \R^q$ of $z_0$ on which
 \eqref{calibration_problem} uniquely defines a function
$\mathcal{O} \ni \pi\mapsto \rho(\pi)$, i.e.\ $\nabla_\varphi cal\mbox{-}err(\pi, \rho (\pi))= 0 $ for all $z \in \mathcal{O} $. Moreover, if  $\rho$ is in  $ \mathcal{C}^1\left(\mathcal{O}; \mathbb{R}^p\right)$, then 
\bel
\partial_{\pi^{\star}} \rho(z) =
- \Big( \partial^2_{\vartheta  ,\vartheta } cal\mbox{-}err(z, \rho(z))\Big)^{-1} \Big( \partial^2_{\vartheta  , \pi }  cal\mbox{-}err(z, \rho(z)) \Big)\sp \mbox{for all $z \in \mathcal{O}$},
\eel
where we assume that 
the matrix of all the second derivatives  of $cal\mbox{-}err$ with respect to one component in $\vartheta$ and one component in $\pi$, denoted by $\partial^2_{\vartheta  , \pi }  cal\mbox{-}err
%_{1 \leq i \leq d}^{ 1 \leq j \leq k}
 $, exists. From there, the chain rule %applied to $\CVA_0( \rho(z))$
\beql{e:chain}
&(\nabla_{\pi} \CVA_0(z))^\top  = \Big(\nabla_{\theta} \CVA_0 ( \rho(z))\Big)^{\top}  \partial_{\pi^{\star}} \rho(z)
\\
&=-\Big(\nabla_{\theta} \CVA_0 ( \rho(z)) \Big)^{\top} \Big( \partial^2_{\vartheta  ,\vartheta } cal\mbox{-}err(z, \rho(z)) \Big)^{-1}  \Big(  \partial^2_{\vartheta  , \pi }  cal\mbox{-}err(z, \rho(z)) \Big)\sp \mbox{for $z\in \mathcal{O}$},
\eeql 
%Replacing $z= z_0$ in \eqref{e:chain} 
allows deducing the market sensitivities $\nabla_{\pi} \CVA_0$ from the model sensitivities $\nabla_{\theta} \CVA_0 $. 
When heavy Monte Carlo (such as CVA) pricing tasks are involved in the $\nabla_{\theta} \CVA_0$ computations,
the time of computing $\nabla_{\pi} \CVA_0$ through \eqref{e:chain} is dominated by the time of computing $\nabla_{\theta} \CVA_0$.

Another way to compute market bump sensitivities is to bump each target calibration price and recompute the price $\Pi_0$ for pricing model parameters $\rho$ recalibrated to each bumped calibrated data set. This direct approach does not need Jacobian transformations and it is also amenable to second-order sensitivities. However, accounting for curves and surfaces of hedging assets, there may be much more market sensitivities than model sensitivities (as in our use case of Section \ref{s:toymo}). Moreover, direct market sensitivities require not only intensive repricings but also model recalibrations, as many as targeted market sensitivities. The direct approach is therefore typically much heavier than the one based on model bump sensitivities followed by Jacobian transformations. We therefore forget direct market bump sensitivities hereafter: by market bump sensitivities, we mean from now on (first-order) bump sensitivities with respect to model parameters, transformed to the corresponding first-order market sensitivities via \eqref{e:chain}.

}

\section{Credit Valuation Adjustment and Its Bump Sensitivities\label{s:thecva}}

    In the above Black-Scholes setup, bump sensitivities are useless because the exact Greeking formulas are also faster.
We now switch to CVA computations in the role of $\Pi$ before, for which pricing and Greeking can only be achieved by intensive Monte Carlo simulations implemented on GPU.
%Figure \ref{fig:speedup}(b). 
Since we are also interested in the risk of CVA fluctuations (if unhedged, or of fluctuations of a hedged CVA position more generally), we now consider the targeted price $\Pi$ (CVA from now on) as a process. 
We denote by $\mathrm{MtM}^c$, the counterparty-risk-free valuation of the portfolio of the bank with its client $c$;
$\tau_c$, the client $c$'s default time, with  intensity process $\gamma^{\thec}$;
 $X$, with $X_0\equiv \bo  $ (componentwise),  the vector of the default indicator processes of the clients of the bank;
$Y$, a
%$\mathbb{R}^{q}$ valued 
diffusive vector process of model risk factors such that each ${\rm MtM}^c_t$ and $\gamma^c_t$ is a measurable function of $(t,Y_t)$, for $t\ge 0$ (in the case of credit derivatives with the client $c$, ${\rm MtM}^c_s$ would also depend on $X_t$, which can be accommodated at no harm in our setup).
The exogenous model parameters are denoted by $\epsilon$.
Let the baseline $\rho_0=(y_0,\epsilon_0)$ denote a calibrated value of $(y,\epsilon)$,
where $y$ is used for referring to the initial condition of $Y$, whenever assumed constant. Let
$\iota$ denote an initial condition for $Y$ randomized around its baseline $y_0$,
$\varepsilon$ 
%$\varepsilon_0$ 
be likewise a randomization of $\epsilon$ around its baseline $\epsilon_0$, 
%inducing the constant-in-time process \b{needed or not?} $\varepsilon \equiv \varepsilon_0$, 
and   
$\varrho_t =(Y_t ,\varepsilon
%_0 
), t\ge 0$.  Starting from the (random) initial condition $(\bo  ,\iota)$, the model $(X,Y)$ is supposed to evolve according to some Markovian dynamics 
 (e.g.\ the one of Section \ref{s:toymo}) parameterized by $\varepsilon
 %_0
 $. 
This setup allows encompassing in a common formalism: 
\begin{itemize}
\item  the \textbf{baseline mode} of  \citet[Section 4]{abbas2021hierarchical} where $\varrho_{0}\equiv\rho_0$;
\item the \textbf{risk mode} where $Y_{0}\equiv y_0$; 
\item the \textbf{sensis mode}, or general $\varrho_{0}$ case, used 
%at $t=0$ 
with an exogenous distribution of $\iota$ in Section
 \ref{s:toymo}
   and with $\iota$ distributed as the diffusive risk factors of the CVA simulation engine in the risk mode 
at a specified risk horizon
   in Section \ref{s:risk}.
\end{itemize}
The CVA engine in the baseline mode $\varrho_{0}\equiv\rho_0$ was introduced in \citet{abbas2021hierarchical}.  
The risk and sensis mode also incorporating a randomization $\varepsilon$ of the exogenous model parameters $\epsilon$, and of $Y_0$ in the sensis mode, are novelties of the present work.

We restrict ourselves to an uncollateralized CVA for notational simplicity. Given $n$ pricing time steps of length $h$ such that $nh=T$, the final maturity of the derivative portfolio of the bank, let, at each $t=ih$,
%,$ $i\in 0\, .. \,n,$
 \beql{eq:LGD} 
 & \LGD_{t}=\sum_c \sum_{j=0}^{ {i-1}}
% {\beta}_{i h}^{-1}{\beta}_{jh}  
(\mathrm{MtM}^c _{jh})^+\indi{jh<\tau_c \leq (j+1)h} ,\\
&
\xi_{t,T} = h
\sum_c  \sum_{j=i}^{{\thenb-1} }  
 ({\mathrm{MtM}}^c_{jh})^+
 (e^{-\sum_{\imath=i}^ {j-1} \gamma^{\thec}_\imath}  - e^{-\sum_{\imath=i}^ {j} \gamma^{\thec}_\imath} )\indi{\left\{\tau_c>i h\right\}.}\eeql
Our computations rely on the following 
default-based and intensity-based 
formulations of the (time-discretized) CVA of a bank with clients $c$, at the pricing time $t=ih$ (cf.\ \citet[Eqns.~(25)-(27)]{abbas2021hierarchical}):
%\footnote{the simulation time $h$ is taken as a divisor of the risk horizon $t$.} \citep*[Eq.~(25) and (27)]{abbas2021hierarchical}:
\beql{eq:CVAint}
& \widetilde{\mathrm{CVA}}_{t} (x,\rho)= \mathbb{E}\Big[
\underbrace{\sum_c\sum_{j=i}^{ {\thenb-1}} {
%\beta}_{i h}^{-1}{\beta}_{jh} 
(\mathrm{MtM}^c _{jh})^+\indi{jh<\tau_c \leq (j+1)h}}}_{\LGD_{T}-\LGD_{t}}
\Big|
X_{ih}=x,\varrho_{ih}=\rho
\Big]  
% \indi{\{\si h<\tau  \} }
%\indi{\{\si h<\tau  \} }, 
\\&\qqq   =%\approx
\mathbb{E}
\Big[\underbrace{h
\sum_c  \sum_{j=i}^{{\thenb-1} }  
 ({\mathrm{MtM}}^c_{jh})^+
 (e^{-\sum_{\imath=i}^ {j-1} \gamma^{\thec}_\imath}  - e^{-\sum_{\imath=i}^ {j} \gamma^{\thec}_\imath} )\indi{\left\{\tau_c>i h\right\}}}  
_{\xi_{t,T}}
\Big|X_{ih}=x,\varrho_{ih}=\rho
\Big],
% \indi{\{\si h<\tau  \} },
\eeql
where each coordinate of $x$ is 0 or 1.
From a numerical viewpoint, the second line of \eqref{eq:CVAint} entails less variance than the first one (see Figure 5 in \citet{abbas2021hierarchical}). Hence we rely for our CVA computations on this second line.
At the initial time 0, as $X_0 \equiv \bo  $, we can restrict attention to the origin $x= \bo  $  and skip the argument $x$ as well as the conditioning by $X_{ih}=x$ in \eqref{eq:CVAint}.
In the sensis mode, one is in the setup
\eqref{e:lepi} for $\xi=\xi_{0,T}$, which implicitly depends on $\varrho_{0}=(\iota,\varepsilon
%_0
)$, i.e.\ $\CVA_0( \varrho_{0}) = \E\big(\xi   (\varrho_{0}  )  \,\big|\,  \varrho_{0} \big) .$
The baseline CVA sensitivities can thus  be 
computed by  
Algorithm \ref{a:fast_sensi},
with $\CVA$ and $(y=Y_0 ,\epsilon)$ here in the role of $\Pi$ and $\rho$ there.
In the baseline mode where $\varrho_{0}\equiv \rho_0$, $\mathrm{CVA}_{0}(\varrho_{0})$ 
is constant, equal to the corresponding \beql{e:cvabaseline} \mathrm{CVA}_{0}(\rho_0)= \E \,\LGD_{nh}=\E\xi_{0,T} ,\eeql which is 
 computed by
 % (or SVD least-squares optimization in the linear case, or 
 Monte Carlo 
 %(no learning needed for a constant) 
 based on the second line of \eqref{eq:CVAint} for $t=0$ there,
as a sample mean of $\xi_{0,T}$, along with the corresponding 95\% confidence interval.

All our CVA computations are done on GPU (whereas the previous Black-Scholes calculations were done on CPU, except for neural net training on GPU). 

\subsection{CVA Lab} \label{s:toymo}

%\subsection{Setup}

In our numerics below, we have 10 economies. For each 
%one 
of them we have a short-term interest rate driven by a Vasicek diffusion and, except for the reference economy, %assumed to be the first out of 10 economies, 
a Black-Scholes exchange rate with respect to the currency of the reference economy. The reference bank has 8 counterparties with corresponding default intensity processes driven by CIR diffusions.
We thus have 8 default indicator processes $X$ of the counterparties and $10+9+8=27$ diffusive risk factors $Y$. This results in a Markovian model $(X,Y)$ of dimension 35, entailing  $p=90$  parameters corresponding to the 27 initial conditions of the $Y$ processes plus their 63 exogenous parameters (see Table \ref{t:sensis} page \pageref{t:sensis}).
This model is only for illustrative purposes: the methodology of the paper can be applied to any Markovian model $(X,Y)$ of client defaults and diffusive (or jump-diffusive if wished) risk factors.
%(with, for instance, time-dependent parameterizations of the long-term mean terms in the Vasicek and CIR diffusions).
% in order to allow for a good fit of the time-0 rate and default intensity curves instead of having them entirely determined, alongside the stochastic dynamics, by a few scalar parameters. We leave these possible extensions to interested
%practitioners on the basis of the provided \b{GitHub} repository.}

A ``reasonably stressed'' but arbitrary baseline $\rho_0$  (see after \eqref{e:d_cva_risk})
%\b{available in the paper's github}
plays the role of calibrated model parameters in our numerics.
In the above pricing model, we consider the CVA on a portfolio of interest rate swaps with characteristics generated randomly  as in \citet[Section 3.3]{abbas2021hierarchical}. 
The portfolio consists of $500$ interest rate swaps with random characteristics (maturity $\le T=10$ years, notional, currency and counterparty) and strikes such that the swaps are worth 0 in the baseline model (i.e.\ for $\varrho_{t}\equiv \rho_0$) at time 0. The swaps 
%(as also the options in the long preprint version\footnote{on \url{https://github.com/hoangdungnguyen/CVA-learning-and-hedging}.\label{f:url}} of the paper) 
have analytic counterparty-risk-free valuation in our pricing model \citep[Section 6]{abbas2021hierarchical}. Their price processes are converted into the reference currency and aggregated
into the corresponding clients MtM$^c$ processes.

We simulate 
by an Euler scheme 
$m = 2^{17}\approx  1.3\times 10^5 $ paths of the pricing model $(X,Y)$,  
with $n = 100$ MtM pricing time steps of length $h=0.1$ and $25$ Euler simulation sub-steps per pricing time step (referred to as daily basis).  
A Monte Carlo computation of \eqref{e:cvabaseline} in the baseline mode then yields $\CVA_0(\rho_0)\in 5,027 \pm 18$ with 95\% probability (computed in about 30s).
A randomization of $Y_0$ and $\epsilon$ in the sensis mode
is used for deriving CVA linear bump sensitivities as per Section \ref{s:fast_ss}. 
Table~\ref{t:sensis} presents benchmark versus fast (linear and \smart) bump sensitivities with respect to the $p=90$ model parameters introduced in Section \ref{s:toymo}. The CIs of the linear and \smart bump sensitivities consistently cover the benchmark bump sensitivities, which, with regard to its much smaller confidence interval, serve as reference for these sensitivities (even if biased with respect to the exact $\partial_\rho\CVA_0$). Regarding the subset simulation trick exposed after Algorithm \ref{a:fast_sensi}, we divided the model parameters into 10 groups separated by  horizontal lines in Table \ref{t:sensis}. 

\newcommand{\padding}{\hspace{-10pt}}
%\definecolor{bronze}{rgb}{0.82, 0.41, 0.12}
\def\orange{BXYZittersweet}\def\orange{orange}
\newcommand{\fp}{\color{\orange}}%{\cellcolor{orange}}
\begin{table}[t!]%[!htbp]
\centering
    \resizebox{\textwidth}{!}{
    \begin{tabular}{|l|rl|rl|rl||l|rl|rl|rl|}
    \hline
param. & \multicolumn{2}{c|}{\begin{tabular}[c]{@{}c@{}}benchmark\\bump\end{tabular}} & \multicolumn{2}{c|}{\begin{tabular}[c]{@{}c@{}}linear\\bump\end{tabular}} & \multicolumn{2}{c||}{\begin{tabular}[c]{@{}c@{}}\smart\\bump\end{tabular}} & param. & \multicolumn{2}{c|}{\begin{tabular}[c]{@{}c@{}}benchmark\\bump\end{tabular}} & \multicolumn{2}{c|}{\begin{tabular}[c]{@{}c@{}}linear\\bump\end{tabular}} & \multicolumn{2}{c|}{\begin{tabular}[c]{@{}c@{}}\smart\\bump \end{tabular}} \\
\hline
$r_0^{\langle 0\rangle}$ & -12,354&\padding$\pm$ 41& -14,426&\padding$\pm$ 1,426& -12,310&\padding$\pm$ 384 & $b^{\langle 8\rangle}$ & -37,295&\padding$\pm$ 487& -43,620&\padding$\pm$ 10,327& -40,903&\padding$\pm$ 4,519 \\
$r_0^{\langle 1\rangle}$ & -4,761&\padding$\pm$ 57& -4,349&\padding$\pm$ 1,331& -4,597&\padding$\pm$ 514 & $b^{\langle 9\rangle}$ & 94,235&\padding$\pm$ 760& 89,363&\padding$\pm$ 11,404& 93,950&\padding$\pm$ 6,802 \\
\cline{8-14}
$r_0^{\langle 2\rangle}$ & 10,715&\padding$\pm$ 92& 12,060&\padding$\pm$ 1,438& 11,010&\padding$\pm$ 859 & \fp$\sigma^{r, \langle 0\rangle}$ &\fp 23,850&\fp\padding$\pm$ 209&\fp 25,610&\fp\padding$\pm$ 5,852&\fp 24,283&\fp\padding$\pm$ 1,979 \\
$r_0^{\langle 3\rangle}$ & 1,433&\padding$\pm$ 37& 1,331&\padding$\pm$ 1,315& 1,521&\padding$\pm$ 353 & \fp$\sigma^{r, \langle 1\rangle}$ &\fp 23,563&\fp\padding$\pm$ 311&\fp 22,360&\fp\padding$\pm$ 4,287&\fp 21,755&\fp\padding$\pm$ 2,460 \\
$r_0^{\langle 4\rangle}$ & 14,712&\padding$\pm$ 62& 14,648&\padding$\pm$ 1,350& 15,054&\padding$\pm$ 573 & \fp$\sigma^{r, \langle 2\rangle}$ &\fp 33,945&\fp\padding$\pm$ 392&\fp 32,240&\fp\padding$\pm$ 4,193&\fp 36,699&\fp\padding$\pm$ 4,635 \\
$r_0^{\langle 5\rangle}$ & 24,539&\padding$\pm$ 146& 26,762&\padding$\pm$ 1,697& 25,057&\padding$\pm$ 1,424 & \fp$\sigma^{r, \langle 3\rangle}$ &\fp 14,402&\fp\padding$\pm$ 191&\fp 16,502&\fp\padding$\pm$ 4,445&\fp 13,434&\fp\padding$\pm$ 1,671 \\
$r_0^{\langle 6\rangle}$ & 15,100&\padding$\pm$ 96& 15,450&\padding$\pm$ 1,416& 14,744&\padding$\pm$ 901 & \fp$\sigma^{r, \langle 4\rangle}$ &\fp 20,347&\fp\padding$\pm$ 292&\fp 18,847&\fp\padding$\pm$ 3,576&\fp 20,866&\fp\padding$\pm$ 2,828 \\
$r_0^{\langle 7\rangle}$ & 29,368&\padding$\pm$ 161& 30,239&\padding$\pm$ 1,612& 29,637&\padding$\pm$ 1,398 & \fp$\sigma^{r, \langle 5\rangle}$ &\fp 36,305&\fp\padding$\pm$ 500&\fp 38,439&\fp\padding$\pm$ 5,455&\fp 34,401&\fp\padding$\pm$ 4,262 \\
$r_0^{\langle 8\rangle}$ & 5,930&\padding$\pm$ 66& 5,410&\padding$\pm$ 1,372& 6,264&\padding$\pm$ 689 & \fp$\sigma^{r, \langle 6\rangle}$ &\fp 26,597&\fp\padding$\pm$ 400&\fp 23,572&\fp\padding$\pm$ 4,182&\fp 26,593&\fp\padding$\pm$ 3,329 \\
$r_0^{\langle 9\rangle}$ & 5,132&\padding$\pm$ 57& 7,117&\padding$\pm$ 1,347& 4,879&\padding$\pm$ 484 & \fp$\sigma^{r, \langle 7\rangle}$ &\fp 31,233&\fp\padding$\pm$ 644&\fp 29,380&\fp\padding$\pm$ 5,086&\fp 32,828&\fp\padding$\pm$ 7,071 \\
\cline{1-7}
$\chi_0^{\langle 1\rangle}$ & 151&\padding$\pm$ 3& 188&\padding$\pm$ 80& 164&\padding$\pm$ 26 & \fp$\sigma^{r, \langle 8\rangle}$ &\fp 28,051&\fp\padding$\pm$ 391&\fp 29,169&\fp\padding$\pm$ 4,154&\fp 25,456&\fp\padding$\pm$ 2,634 \\
$\chi_0^{\langle 2\rangle}$ & 733&\padding$\pm$ 7& 709&\padding$\pm$ 86& 757&\padding$\pm$ 78 & \fp$\sigma^{r, \langle 9\rangle}$ &\fp 24,085&\fp\padding$\pm$ 322&\fp 22,284&\fp\padding$\pm$ 3,722&\fp 24,944&\fp\padding$\pm$ 3,326 \\
\cline{8-14}
$\chi_0^{\langle 3\rangle}$ & 123&\padding$\pm$ 2& 134&\padding$\pm$ 77& 134&\padding$\pm$ 21 & \fp$\sigma^{\chi,\langle 1\rangle}$ &\fp 292&\fp\padding$\pm$ 10&\fp 194&\fp\padding$\pm$ 211&\fp 364&\fp\padding$\pm$ 103 \\
$\chi_0^{\langle 4\rangle}$ & 816&\padding$\pm$ 6& 790&\padding$\pm$ 81& 808&\padding$\pm$ 56 & \fp$\sigma^{\chi,\langle 2\rangle}$ &\fp 406&\fp\padding$\pm$ 21&\fp 317&\fp\padding$\pm$ 198&\fp 378&\fp\padding$\pm$ 181 \\
$\chi_0^{\langle 5\rangle}$ & 829&\padding$\pm$ 8& 932&\padding$\pm$ 85& 941&\padding$\pm$ 83 & \fp$\sigma^{\chi,\langle 3\rangle}$ &\fp 224&\fp\padding$\pm$ 8&\fp 229&\fp\padding$\pm$ 189&\fp 185&\fp\padding$\pm$ 49 \\
$\chi_0^{\langle 6\rangle}$ & 835&\padding$\pm$ 9& 861&\padding$\pm$ 96& 852&\padding$\pm$ 89 & \fp$\sigma^{\chi,\langle 4\rangle}$ &\fp 300&\fp\padding$\pm$ 18&\fp -31&\fp\padding$\pm$ 251&\fp 247&\fp\padding$\pm$ 139 \\
$\chi_0^{\langle 7\rangle}$ & 1,030&\padding$\pm$ 11& 1,036&\padding$\pm$ 92& 979&\padding$\pm$ 92 & \fp$\sigma^{\chi,\langle 5\rangle}$ &\fp 460&\fp\padding$\pm$ 23&\fp 426&\fp\padding$\pm$ 205&\fp 526&\fp\padding$\pm$ 233 \\
$\chi_0^{\langle 8\rangle}$ & 243&\padding$\pm$ 4& 213&\padding$\pm$ 77& 243&\padding$\pm$ 32 & \fp$\sigma^{\chi,\langle 6\rangle}$ &\fp 543&\fp\padding$\pm$ 29&\fp 440&\fp\padding$\pm$ 200&\fp 487&\fp\padding$\pm$ 220 \\
$\chi_0^{\langle 9\rangle}$ & 583&\padding$\pm$ 6& 543&\padding$\pm$ 85& 559&\padding$\pm$ 50 & \fp$\sigma^{\chi,\langle 7\rangle}$ &\fp 458&\fp\padding$\pm$ 36&\fp 394&\fp\padding$\pm$ 229&\fp 211&\fp\padding$\pm$ 191 \\
\cline{1-7}
$\gamma_0^{\langle 1\rangle}$ & 2,201&\padding$\pm$ 15& 2,371&\padding$\pm$ 266& 2,155&\padding$\pm$ 131 & \fp$\sigma^{\chi,\langle 8\rangle}$ &\fp 402&\fp\padding$\pm$ 13&\fp 405&\fp\padding$\pm$ 210&\fp 307&\fp\padding$\pm$ 66 \\
$\gamma_0^{\langle 2\rangle}$ & 1,528&\padding$\pm$ 12& 1,554&\padding$\pm$ 261& 1,498&\padding$\pm$ 109 & \fp$\sigma^{\chi,\langle 9\rangle}$ &\fp 344&\fp\padding$\pm$ 20&\fp 401&\fp\padding$\pm$ 214&\fp 459&\fp\padding$\pm$ 200 \\
\cline{8-14}
$\gamma_0^{\langle 3\rangle}$ & 3,097&\padding$\pm$ 24& 2,843&\padding$\pm$ 267& 3,133&\padding$\pm$ 228 & $\alpha_0^{\langle 1\rangle}$ & 86&\padding$\pm$ 1& 86&\padding$\pm$ 19& 82&\padding$\pm$ 11 \\
$\gamma_0^{\langle 4\rangle}$ & 1,250&\padding$\pm$ 10& 1,447&\padding$\pm$ 255& 1,280&\padding$\pm$ 92 & $\alpha_0^{\langle 2\rangle}$ & 69&\padding$\pm$ 1& 66&\padding$\pm$ 19& 68&\padding$\pm$ 9 \\
$\gamma_0^{\langle 5\rangle}$ & 1,473&\padding$\pm$ 12& 1,466&\padding$\pm$ 263& 1,384&\padding$\pm$ 103 & $\alpha_0^{\langle 3\rangle}$ & 143&\padding$\pm$ 2& 140&\padding$\pm$ 21& 152&\padding$\pm$ 19 \\
$\gamma_0^{\langle 6\rangle}$ & 2,982&\padding$\pm$ 15& 2,964&\padding$\pm$ 276& 2,937&\padding$\pm$ 136 & $\alpha_0^{\langle 4\rangle}$ & 38&\padding$\pm$ 1& 45&\padding$\pm$ 14& 41&\padding$\pm$ 5 \\
$\gamma_0^{\langle 7\rangle}$ & 6,068&\padding$\pm$ 32& 6,122&\padding$\pm$ 321& 6,001&\padding$\pm$ 306 & $\alpha_0^{\langle 5\rangle}$ & 45&\padding$\pm$ 1& 58&\padding$\pm$ 18& 48&\padding$\pm$ 6 \\
$\gamma_0^{\langle 8\rangle}$ & 5,887&\padding$\pm$ 27& 5,976&\padding$\pm$ 310& 5,976&\padding$\pm$ 258 & $\alpha_0^{\langle 6\rangle}$ & 154&\padding$\pm$ 1& 158&\padding$\pm$ 18& 165&\padding$\pm$ 11 \\
\cline{1-7}
$a^{\langle 0\rangle}$ & -1,125&\padding$\pm$ 5& -1,142&\padding$\pm$ 91& -1,116&\padding$\pm$ 47 & $\alpha_0^{\langle 7\rangle}$ & 336&\padding$\pm$ 3& 336&\padding$\pm$ 23& 315&\padding$\pm$ 28 \\
$a^{\langle 1\rangle}$ & -823&\padding$\pm$ 10& -831&\padding$\pm$ 103& -838&\padding$\pm$ 83 & $\alpha_0^{\langle 8\rangle}$ & 285&\padding$\pm$ 2& 283&\padding$\pm$ 19& 270&\padding$\pm$ 25 \\
\cline{8-14}
$a^{\langle 2\rangle}$ & 133&\padding$\pm$ 9& 162&\padding$\pm$ 102& 120&\padding$\pm$ 92 & $\delta^{\langle 1\rangle}$ & 6,386&\padding$\pm$ 53& 5,818&\padding$\pm$ 1,185& 6,032&\padding$\pm$ 437 \\
$a^{\langle 3\rangle}$ & -240&\padding$\pm$ 4& -249&\padding$\pm$ 83& -239&\padding$\pm$ 38 & $\delta^{\langle 2\rangle}$ & 6,737&\padding$\pm$ 53& 6,831&\padding$\pm$ 1,236& 6,556&\padding$\pm$ 440 \\
$a^{\langle 4\rangle}$ & 570&\padding$\pm$ 7& 506&\padding$\pm$ 91& 519&\padding$\pm$ 72 & $\delta^{\langle 3\rangle}$ & 8,693&\padding$\pm$ 91& 9,617&\padding$\pm$ 1,212& 8,476&\padding$\pm$ 738 \\
$a^{\langle 5\rangle}$ & 1,093&\padding$\pm$ 11& 1,055&\padding$\pm$ 123& 1,097&\padding$\pm$ 98 & $\delta^{\langle 4\rangle}$ & 6,096&\padding$\pm$ 42& 6,372&\padding$\pm$ 1,183& 5,928&\padding$\pm$ 358 \\
$a^{\langle 6\rangle}$ & 660&\padding$\pm$ 9& 681&\padding$\pm$ 95& 728&\padding$\pm$ 81 & $\delta^{\langle 5\rangle}$ & 5,888&\padding$\pm$ 36& 5,708&\padding$\pm$ 1,276& 5,846&\padding$\pm$ 305 \\
$a^{\langle 7\rangle}$ & 1,377&\padding$\pm$ 13& 1,462&\padding$\pm$ 118& 1,421&\padding$\pm$ 117 & $\delta^{\langle 6\rangle}$ & 14,539&\padding$\pm$ 67& 14,824&\padding$\pm$ 1,185& 15,038&\padding$\pm$ 686 \\
$a^{\langle 8\rangle}$ & -482&\padding$\pm$ 11& -430&\padding$\pm$ 112& -509&\padding$\pm$ 91 & $\delta^{\langle 7\rangle}$ & 23,261&\padding$\pm$ 128& 21,714&\padding$\pm$ 1,336& 23,014&\padding$\pm$ 1,082 \\
$a^{\langle 9\rangle}$ & -68&\padding$\pm$ 7& -56&\padding$\pm$ 106& -69&\padding$\pm$ 73 & $\delta^{\langle 8\rangle}$ & 31,441&\padding$\pm$ 144& 31,949&\padding$\pm$ 1,484& 31,938&\padding$\pm$ 1,558 \\
\cline{1-7}
\cline{8-14}
$b^{\langle 0\rangle}$ & -166,788&\padding$\pm$ 437& -169,247&\padding$\pm$ 10,994& -166,618&\padding$\pm$ 4,384 & \fp$\nu^{\langle 1\rangle}$ &\fp -38&\fp\padding$\pm$ 8&\fp -78&\fp\padding$\pm$ 132&\fp 16&\fp\padding$\pm$ 84 \\
$b^{\langle 1\rangle}$ & -31,802&\padding$\pm$ 406& -38,577&\padding$\pm$ 9,149& -31,678&\padding$\pm$ 4,753 & \fp$\nu^{\langle 2\rangle}$ &\fp -47&\fp\padding$\pm$ 8&\fp -53&\fp\padding$\pm$ 132&\fp -147&\fp\padding$\pm$ 59 \\
$b^{\langle 2\rangle}$ & 78,709&\padding$\pm$ 823& 81,995&\padding$\pm$ 10,636& 86,000&\padding$\pm$ 8,440 & \fp$\nu^{\langle 3\rangle}$ &\fp -57&\fp\padding$\pm$ 15&\fp -87&\fp\padding$\pm$ 138&\fp 6&\fp\padding$\pm$ 132 \\
$b^{\langle 3\rangle}$ & -6,206&\padding$\pm$ 341& -2,364&\padding$\pm$ 10,461& -3,459&\padding$\pm$ 3,561 & \fp$\nu^{\langle 4\rangle}$ &\fp -26&\fp\padding$\pm$ 6&\fp -10&\fp\padding$\pm$ 126&\fp -37&\fp\padding$\pm$ 62 \\
$b^{\langle 4\rangle}$ & 140,127&\padding$\pm$ 683& 150,195&\padding$\pm$ 10,716& 138,407&\padding$\pm$ 6,189 & \fp$\nu^{\langle 5\rangle}$ &\fp -35&\fp\padding$\pm$ 6&\fp 45&\fp\padding$\pm$ 128&\fp -44&\fp\padding$\pm$ 49 \\
$b^{\langle 5\rangle}$ & 114,437&\padding$\pm$ 914& 109,486&\padding$\pm$ 10,779& 121,754&\padding$\pm$ 9,471 & \fp$\nu^{\langle 6\rangle}$ &\fp -66&\fp\padding$\pm$ 13&\fp -60&\fp\padding$\pm$ 151&\fp -100&\fp\padding$\pm$ 118 \\
$b^{\langle 6\rangle}$ & 127,783&\padding$\pm$ 1,108& 120,891&\padding$\pm$ 9,923& 123,214&\padding$\pm$ 9,430 & \fp$\nu^{\langle 7\rangle}$ &\fp -151&\fp\padding$\pm$ 23&\fp -244&\fp\padding$\pm$ 146&\fp -150&\fp\padding$\pm$ 209 \\
$b^{\langle 7\rangle}$ & 191,031&\padding$\pm$ 1,373& 188,073&\padding$\pm$ 12,629& 196,130&\padding$\pm$ 14,048 & \fp$\nu^{\langle 8\rangle}$ &\fp -161&\fp\padding$\pm$ 24&\fp -230&\fp\padding$\pm$ 154&\fp -129&\fp\padding$\pm$ 221 \\
\hline
\hline
    \end{tabular}
}
\caption{CVA sensitivities and 95\% confidence intervals 
(CI)
with respect to the $p=90$ parameters of the model, estimated by the benchmark, linear and \smart bump approaches.  
%benchmark bump (i.e.\ \pointwise bump using $m$ paths per sensitivity), \smart bump (i.e.\ \pointwise bump using $m/p$ paths per sensitivity), and linear bump (using a common simulation run with $m$ paths for all the sensitivities) approaches. 
The notation for the parameters follows \citet[Eqns.\ (43) and (44)]{abbas2021hierarchical}. 
The horizontal lines separate groups of parameters bumped simultaneously under the linear bump approach, the way explained after Algorithm \ref{a:fast_sensi}. The model volatility parameters left aside from the calibration process (see the comments to Figure \ref{fig:zc_cds_sensis} below) are in {\color{\orange}{orange}}.\label{t:sensis}}
\end{table}

\begin{figure}[t]
% \resizebox{\textwidth}{!}
% %{
%\vspace*{-1cm}
%\hspace*{-1cm}
\begin{subfigure}{1.0\linewidth}
\hspace{-0.cm}\includegraphics[width=\linewidth]{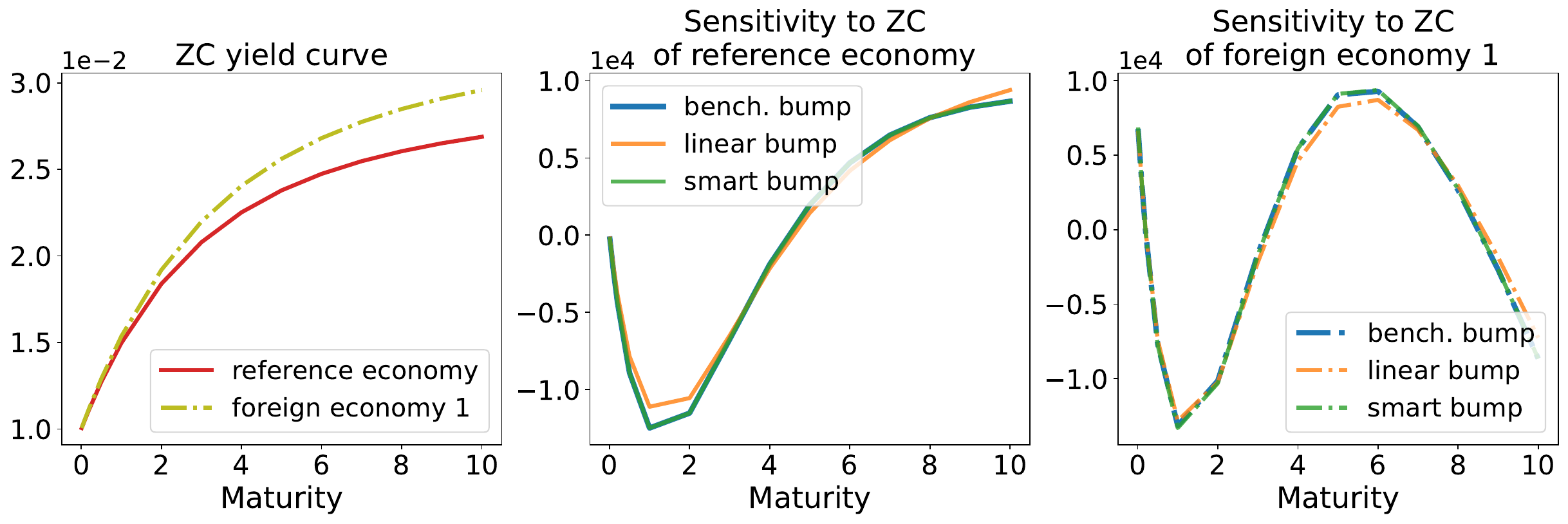}
\end{subfigure}
\begin{subfigure}{1.0\linewidth}
\hspace{-0.cm}\includegraphics[width=\linewidth]{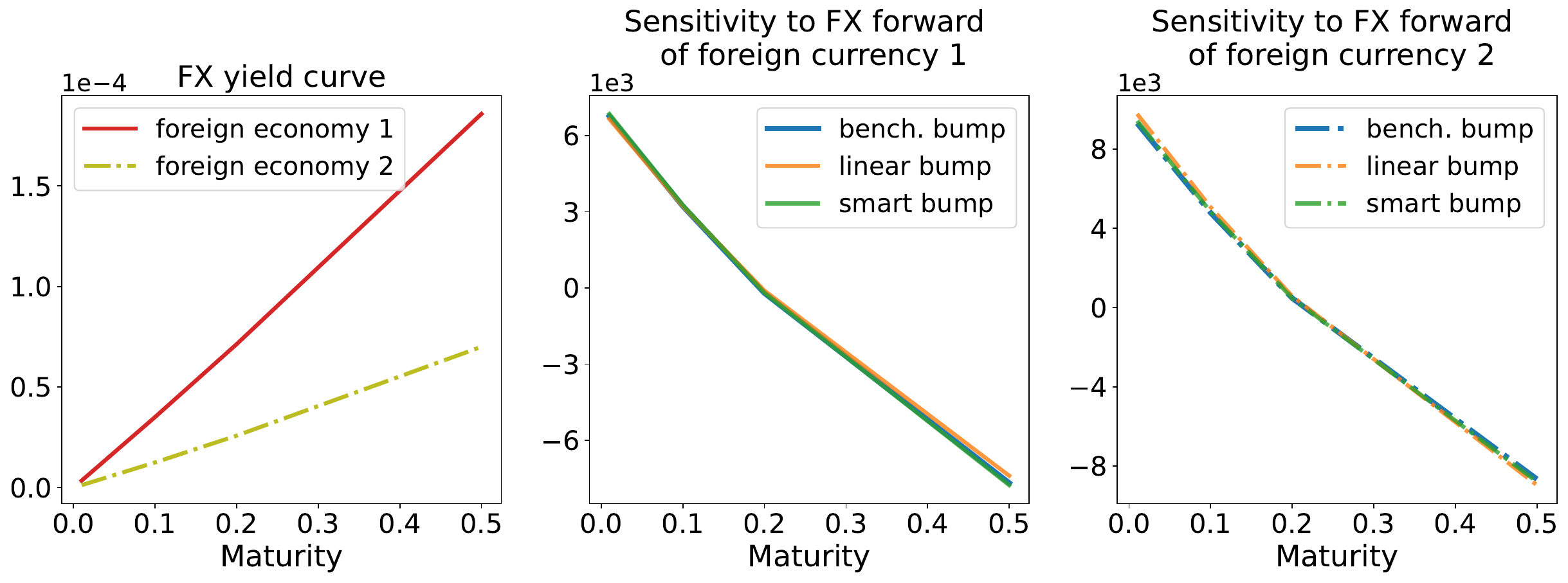}
\end{subfigure}

%\begin{subfigure}{1.0\linewidth}
%\hspace{-0.cm}\includegraphics[width=\linewidth]{new_fig/cds_sensi.pdf}
%\end{subfigure}
\begin{subfigure}{1.0\linewidth}
\hspace{-0.cm}\includegraphics[width=\linewidth]{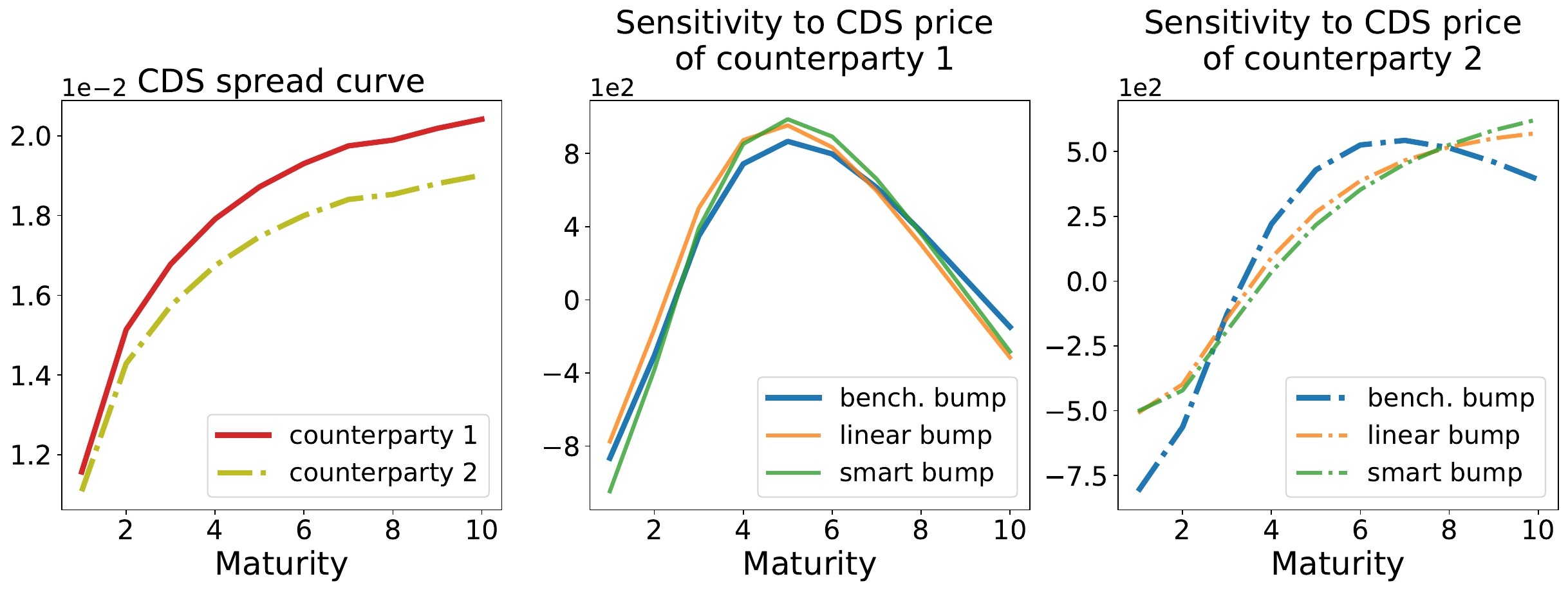}
\end{subfigure}

 \caption{ Sensitivities to prices of selected zero-coupons (ZCs) ({\textit{top}}), FX forward contracts  %\r{change curves and titles to 
 %FX forward term structure}
 %\b{also, this FX curve should just be the expected fx spot term structure} 
 (\textit{middle}), and CDS contracts (\textit{bottom}), estimated by Jacobian transformations from benchmark and fast (linear and \smart) bump sensitivities to the pricing model parameters. ZC and FX yield curves are constructed  from ZC and FX forward prices through the transformation $\cdot \mapsto -\frac{\log(\cdot)}{T}$, where $T$ denotes the maturity.}
\label{fig:zc_cds_sensis}
\end{figure}

 For converting the model sensitivities into market sensitivities by the Jacobian method of Section  \ref{ss:jac}, we first need to specify the market instruments. In this case, each of the 10 zero-yield curves has the 14 pillars $ 0.01, 0.1, 0.2, 0.5, 1 ,\dots, 10$ years, each of the 9 FX forward curves has 4 pillars $ 0.01, 0.1, 0.2, 0.5$ year, and each of the 8 CDS curves (with monthly payments and loss-given-default parameter set to 60\% for each counterparty) has the 10 pillars $1 ,\dots, 10$ years, resulting in a total of $q=256$
 % \b{notation $d$ conflicting with BS: problem or not? remove $d$ there? make $d$ $q$ here, but $q$ also quantile?} 
 market instruments and first-order market sensitivities.
As
 %for simplicity 
 we do not introduce any options as hedging
 %/ calibration
 assets, %For simplicity \b{1) necessary even in the \pointwise bump case? 2) can one really 'freeze then randomize'?}, 
we freeze the volatility model parameters (in orange in Table \ref{t:sensis}) and only consider the calibration error 
 %$cal\mbox{-}err$ 
 as a function of the initial conditions and drift parameters of the model risk factors.
The reason why our FX curves are restricted to 4 points 
%on the short end of the curve 
is because we thus only have one FX-related model (spot exchange rate) parameter for each foreign currency.
 Figure \ref{fig:zc_cds_sensis} page \pageref{fig:zc_cds_sensis} displays some of the 
interest-rate, credit and FX
market sensitivities deduced from the model sensitivities of Table \ref{t:sensis} by Jacobian transformation the way explained in Section \ref{ss:jac}. We note the consistency between the orders and magnitudes, but also term structure profiles (except in the ends of the second credit curve), of the market sensitivities deduced from the benchmark vs.\ fast (linear and smart) bump sensitivities, with again a slight advantage of \smart over linear bump sensitivities.

\begin{figure}[t] 
\begin{subfigure}{\linewidth}\centering
\includegraphics[width=\linewidth]{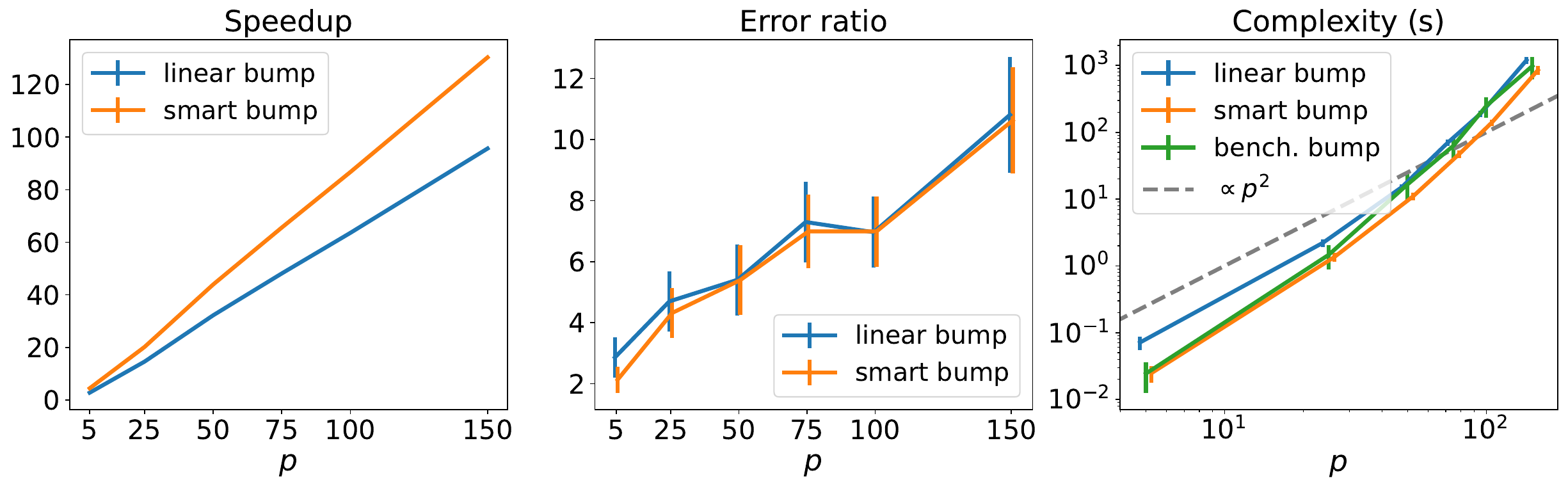}
\vspace{-20pt}
\caption{Performance in Black-Scholes computations}
\label{fig:perf_BS}
\end{subfigure}
\begin{subfigure}{
\linewidth}\centering
\includegraphics[width=\linewidth]{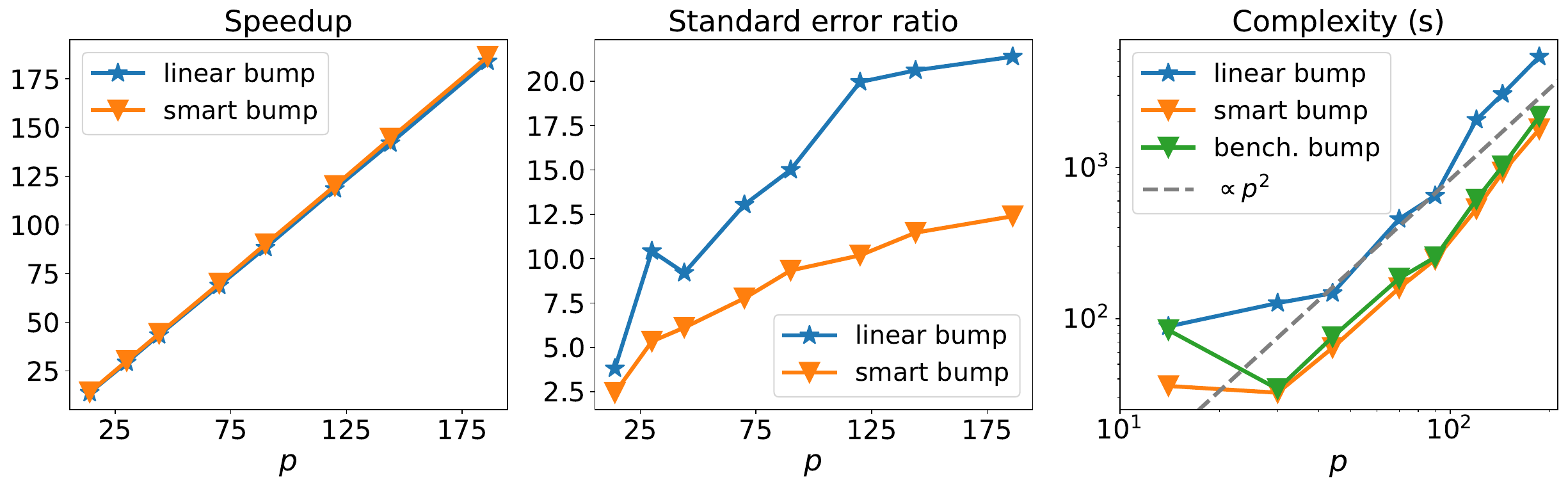}
\vspace{-20pt}
\caption{Performance in CVA computations}
\label{fig:perf_CVA}
\end{subfigure}
\vspace*{-0.1cm}
\caption{ Comparison of performance of linear and \smart bump sensitivities as a function of the number of model parameters $p$ fixing $m=2^{16}$ in (a) and $m=2^{17}$ in (b), using benchmark bump sensitivities as references. 
The speedups displayed in the left panels are obtained by dividing the execution times for benchmark sensitivities by the execution times for linear or \smart %\r{model} 
bump sensitivities.
% \b{For each $p$, the notion of error reported in the middle panels is: in the Black-Scholes case,
% the mean of the $p$ differences between true sensitivities and bump sensitivities divided by the true sensitivities; in the CVA case,
% the median of the $p$ Monte Carlo relative standard errors, where each Monte Carlo relative standard error is obtained as the Monte Carlo standard error of the linear or \smart bump sensitivities divided by the absolute value of the benchmark sensitivities.}
For each value of $p$, the errors displayed in the middle panel are:
in the Black-Scholes case, the  mean 
 of the $p$ ratios between
the
relative error of the linear (resp.\ \smart) bump sensitivity and the relative error of the benchmark sensitivity; in the CVA case, the median of the $p$ ratios between the Monte Carlo standard errors of the linear (resp.\ \smart) bump sensitivity and of the benchmark sensitivity.
 The complexities in the right panels are defined as the projected times of reaching a  relative error (for Black-Scholes) or relative standard error (for CVA) of $1\%$
 given the $1/\sqrt{| \mbox{sample size} |}$ scaling of the error and standard error, with $| \mbox{sample size} |=m$ for linear and benchmark bumps and $m/p$ for \smart bumps;
that is,
$\mbox{projected time (i.e.\ complexity)}=\mbox{exec.\ time}\times 
10^4 \mbox{ \big(relative (standard) error\big)}^2 $.
The $\sqrt{| \mbox{sample size} |}$ above is justified as explained after \eqref{e:theta} for linear bumps and by the central limit theorem for benchmark and \smart bumps. 
The performance metrics of the Black-Scholes computations rely on $64$ Monte Carlo simulation runs per $p$,  with associated confidence intervals computed according to the principle of uncertainty propagation in \citep[Chapter 4 page 22]{lee2005analyzing}.
The performances of \CVA sensitivities are produced from a single Monte Carlo simulations run per $p$ (as for a given run we only have access to CVA sensitivities errors in the sense of standard errors, which cannot be aggregated across different runs).}

\label{fig:speedup}
\end{figure}

The times taken to generate the model sensitivities of Table~\ref{t:sensis} and the corresponding market sensitivities are recorded in Table \ref{tab:timesensis}.
The linear and \smart bump calculations 
%and the \smart regression calculations 
use the same $m$ Brownian driving paths
% use the same number $m$ of Brownian driving paths 
$\omega$ of the pricing model as the one used for the Monte Carlo computation of $\CVA_0(\rho_0)$.
%Each \smart bump sensitivity uses $m / p$ paths of a Monte Carlo simulation run with $m$ paths as a whole (which is significantly more GPU efficient to simulate the \CVA payoffs than doing $p$ Monte Carlo runs of size $m / p$ each), while each benchmark bump sensitivity uses the totality of the $m$ paths.
%The linear bump sensitivities achieve a \r{34.9} times speedup compared to the benchmark ones. 
%Due to the common overhead of estimating Jacobian matrix, the speedup drops to \r{16.0} for market sensitivities. %The linear bump sensitivities are 4.4 times faster than the \smart bump sensitivities, while achieving similar performance in terms of error as shown in Figure~\ref{fig:speedup}(b).
The speedups of the linear and \smart bump sensitivities are %even more significant
almost identical, of about 90 with respect to the benchmark bump sensitivities. %\r{The times shown in Table~\ref{tab:timesensis} exclude the avoidable repeated overheads for pricing.}%and \r{15.9} times for the market sensitivities compared with the benchmark sensitivities. %as shown in 
 In Figure~\ref{fig:speedup}(b) page \pageref{fig:speedup}, 
%On this figure 
the increasing curves in the left panel highlight the almost linear growth of the speedup of the linear  and \smart bump sensitivities with respect to the benchmark ones when the number $p$ of pricing model parameters increases, but with also increasing errors displayed in the middle panel.  
The \smart bump sensitivities have smaller confidence interval than the linear bump sensitivities for all tested $p$. Combining with the timing result, we conclude that the \smart bump sensitivities outperform the linear bump sensitivities in this \CVA use case. This is confirmed by the right panel of Figure~\ref{fig:speedup}(b), where for all tested $p$ the complexity of linear bump sensitivities is  higher than the one of \smart bump sensitivities, itself very close (as expected) to %(slightly better than) 
the one of the benchmark bump sensitivities.
  \begin{table}[H]%[!ht]
    \centering
    %\rule{6.4cm}{3.6cm}
    %
    \resizebox{0.9\textwidth}{!}{
    \begin{tabular}[b]{|l|c|c|c|c|c|}
        \hline
        \multirow{3}{*}{} & \multicolumn{5}{c|}{Market sensitivities}\\
        \cline{2-6}
         & \multicolumn{3}{c|}{Parameter sensitivities} & \multirow{2}{*}{Jacobian} & \multirow{2}{*}{Total} \\
         \cline{2-4}
         & Simul. & Lin. regr. & Total & & \\
        \hline
        %bench. & 13m47s & N/A & 13m47s(1$\times$) & \multirow{3}{*}{30s} & 14m17s(1$\times$)\\
        %\cline{1-4} \cline{6-6}
        %linear & 23.7s & 0.1s & 23.7s(34.9$\times$)  &  & 53.7s(16.0$\times$) \\
        %\cline{1-4} \cline{6-6}
        %\smart & 23.9s & N/A & 23.9s(34.6$\times$)  &  & 53.9s(15.9$\times$) \\
        bench. & 12m48s & N/A & 12m48s(1$\times$) & \multirow{3}{*}{30s} & 13m18s(1$\times$)\\
        \cline{1-4} \cline{6-6}
        linear & 8.6s & 0.1s & 8.7s(88.5$\times$)  &  & 38.7s(20.6$\times$) \\
        \cline{1-4} \cline{6-6}
        \smart & 8.5s & N/A & 8.5s(90.0$\times$)  &  & 38.5s(20.7$\times$) \\
        \hline
        \end{tabular}
    }
    %\captionsetup{labelformat=table}
    \caption{Computation times (and speedups ``$\times$'' in parentheses) for CVA bump sensitivities shown in Table~\ref{t:sensis}. The time of the benchmark bump sensitivities contains diagonal gamma estimation (needed for the ``bench.\ bump sensis w/ $\Gamma$''
    %some of the CVA run-on risk proxies of
    in Table \ref{t:sensi} page \pageref{t:sensi}), which amounts to an extra 5 seconds needed for computing the baseline $\CVA_0(\rho_0)$
    (the $\CVA_0(\varrho_{0})$ and $\CVA_0(\varrhob_0)$ being already needed for the deltas). Here and in Figure~\ref{fig:speedup}(b) page \pageref{fig:speedup} likewise, the times exclude the infrastructure initialization times taken once for all computations (allocating GPU memory, compiling CUDA kernel, etc.), which amount to $10$ to $30$ seconds depending on $p$.
}\label{tab:timesensis}
  \end{table}

\FloatBarrier

\section{Learning the Future CVA}\label{s:cvalh}

Equivalently to the second line in \eqref{eq:CVAint},
%(assuming $\xi_{t,T}$ square integrable)
%$\widetilde{\mathrm{CVA}}_{ih}= \tilde{\varphi
%} 
%_{i} 
%(X_{ih},Y_{ih}) $ with 
\beql{e:onecliint}
&\mathrm{CVA}_{t}(\cdot)
 %\tilde{\varphi}_{i}
%^{\thec} 
= \argmin_{\Phi \in\cB }
%\\&\qqq
\mathbb{E}
  \Big[ \big(
  \xi_{t,T}
%   \sum_c \sum_{j=i}^{{\thenb-1} }  
% h  ({\mathrm{MtM}}^c_{jh})^+
% (e^{-\sum_{\imath=i}^ {j-1} \gamma^{\thec}_\imath}-e^{-\sum_{\imath=i}^ {j
% } \gamma^{\thec}_\imath})
%  \indi{\left\{\tau_c>i h\right\}}
-\Phi(X_{t},\varrho_{t}) \big)^2  \Big]
,\eeql  
where
$\cB$
%( {\mathbb{R}^{q}} )$ 
is the set of the Borel measurable functions of $(x,\rho)$. 
We denote by ${\CVA}_{\theh}^{\theta}(X_t,\varrho_t)$ the conditional \CVA at time 
$\theh=ih> 0$ learned by a neural network 
with parameters $\theta$ on the basis of simulated pairs $(X_t,\varrho_t)$ and cash flows $\xi_{t,T}$. The conditional CVA pricing function ${\CVA}_{\theh}^{\theta}(x,\rho)$ is obtained by
replacing $\mathbb{E}$ by a simulated sample mean $\widehat{\mathbb{E}}$
% (also used for computing CVA$_0$ based on the second line of \eqref{eq:CVAint} for $t=0$ there) 
  and $\cB
  %( {\mathbb{R}^{q}} )
  $ by 
 a 
 %(possibly linear) 
 neural net (or linear as a special case) search space 
 %$\mathcal{NN}$ 
 in the optimization problem 
\eqref{e:onecliint}.
%\footnote{in the Bayesian-mixture variant of the approach, model parameters (which are implicit in \eqref{eq:CVAint})  randomly drawn at time 0 are added to the regressors (inputs of the neural net) $(X_{ih} ,Y_{ih} )$, the way detailed in Section \ref{s:bayes}.}.
The latter is then addressed  numerically by Adam mini-batch  gradient descent on the basis of simulated pairs $(X_t,\varrho_t)$ as features and $\xi_{t,T}$ as labels:  
see Algorithm \ref{alg:cva_learning}, 
in the baseline and risk modes. 

\begin{algorithm}[H] 
\small
\LinesNumbered
\SetAlgoLined
\SetKwInOut{AlgName}{name}
\SetKwInOut{Input}{input}\SetKwInOut{Output}{output} 
\Input{Calibrated pricing model parameters $\rho_0 = (y_0, \epsilon_0)$ (with client default indicators $X_0$ all equal to 0), 
a CVA pricing time $t=ih >0$, a number of paths $m$ with $n$ pricing time steps (and daily Euler simulation step) of the pricing model, a number of training epochs $E$ (e.g. 1000) of the (neural network or linear) learning model for $\CVA_t(X_t, \varrho_t)$,  a  partition $B$ of $\{1,\dots, m\}$ into mini-batches. Adam optimizer set by default.
}

\Output{one trained set $\theta$ of the parameters of the predictor  $\CVA^\theta_t(x,y)\approx \CVA_t(x,y)$ in the baseline mode or $\CVA^\theta_t(x,y, \epsilon)\approx \CVA_t(x,y, \epsilon)$ in the risk mode. 
} 

\uIf{baseline mode}{
Set $\epsilon_j = \epsilon_{0}$, for $j = 1\,..\,m$
}\ElseIf{risk mode}{
Draw $m$ i.i.d. bumped exogenous model parameters $  \epsilon_j $ from the distribution  $\mathcal{N} \Big( \epsilon_0, \Id\big({\sigma^2}\epsilon_0 \odot \epsilon_0\big)  \Big)$  
%taking $X$ and $Y$ as input
}
For each $j\in 1\,..\,m$ with associated $\rho_j=(y_0,\epsilon_j)$ and driving Brownian path $\omega_j$, simulate one pricing model path $(X,Y)(\rho_j;\rmo _j)$ 
%with $n$ time steps 
starting from $(\bo  , y_0)$ with exogenous model parameters $\epsilon_j$ and compute the corresponding $\xi_{t,T}(\rho_j;\rmo _j)$ 
%paths ${\mathrm{MtM}}^c,\gamma^c$ and time $\tau_c$ for each client $c$ of the bank

Initialize a neural network $%\Phi 
\CVA^\theta_t(x,y) 
%\in \mathcal{NN}
$ in the baseline mode
or $%\Phi 
\CVA^\theta_t(x,y,\epsilon) $
in the risk mode

Define the loss function $\mathcal{L}(\theta, batch) = $ sample mean over trajectories in $batch \in B$ of
the (cf.\ \eqref{e:onecliint}) $\Big(
  \xi_{t,T} (\rho_j ;\rmo _j)
-\Phi^\theta_j 
\Big)^2$, with 
$\Phi^\theta_j=\CVA^\theta_t\Big(X_t(\rho_j;\omega_j ),Y_t(\rho_j;\omega_j   \Big)$ in the baseline mode
 or
 $\CVA^\theta_t\Big(X_t(\rho_j; \omega_j),Y_t(\rho_j; \omega_j),\epsilon_j\Big)$ in the risk mode
\\
 \For{$\text{epoch} = 1, \dots, E$}{
  \For{$\text{batch} \in B$}{
$\theta\leftarrow \text{AdamStep}(\mathcal{L}(\theta, batch) )$\\
 }
 }
  %\SetKwFunction{Ret}{\bf return}  \Ret{bla}
%Simulate $m$ new (independent) model scenarios $(X_{t},Y_{t})$ (for out-of-sample prediction) and compute 
%$ 
%    \CVA^{\theta}_t = \Phi _i^\theta(X_{t},Y_{t}) 
%$ on each out-of-sample scenario.
    
    \caption{Learning CVA$_t(X_t,\varrho_t)$ at some future time $t> 0$ in the baseline or risk mode. 
The  distribution of $\varepsilon$ in the risk mode can be set exogenously, as displayed above for simplicity, or assessed statistically on a historical basis, as done in the SIMM context of \citeN{albanese2017var}.
    }
\label{alg:cva_learning}
\end{algorithm}

\FloatBarrier

% using an appropriate
% %(depending on the application at hand, compare e.g.\ Sections \ref{s:toymo} and \ref{s:cpes}) 
% distribution for simulating the $\epsilon_j
% $ around their baseline $\epsilon_0$, as well as also various possible initial conditions $y_j$ of $Y$ around their baseline $y_{0}$. 
% Each draw of model parameters $\rho_j=(y_j,\epsilon_j)$
% and its symmetric $\rhob_j=2\rho_0- \rho_j$  are then used (along with $X_0\equiv 0$) for simulating two trajectories 
% %$(X,Y)(\rho_j;\omega_j)$ and $(X,Y)(\bar{\rho}_j;\omega_j)$
% of the pricing model driven by the same Brownian paths $\omega_j$, with resulting cash flows 
% %$(X,Y)(\rho_j;\omega_j)$ and $(X,Y)(\bar{\rho}_j;\omega_j)$
% $\xi(\rho_j;\omega_j)$ and $\xi(\bar{\rho}_j;\omega_j)$. 
% The CVA linear bump sensitivities are then regressed from these simulated data 
% the way detailed in
% Algorithm \ref{a:fast_sensi} (see e.g.~Table~\ref{t:sensis}). CVA \smart bump sensitivities are computed similarly, with linear regression tantamount in their case to local sample means over $m/p$ paths.

%\end{table}
%\FloatBarrier
 \subsection{Twin Monte Carlo Validation Procedure\label{ss:twin}}

\def\Theta{\Pi}
\def\chi{\xi}
A key asset of probabilistic machine learning procedures for any conditional expectation such as $\Pi_0( \varrho   )$ in Section 
\ref{s:fs} %, whose notations we adopt again in this part,
%\eqref{e:learnpi} or \eqref{e:theta},
is the availability of the companion ``twin Monte Carlo validation procedure'' of \citet[Section 2.4]{abbas2021hierarchical}, allowing one to assess the accuracy of a predictor. 
Let $\chi^{(1)} (\varrho ) $
%$\omega\mapsto \chi^{(1)}  (\rho ;\omega) $ 
and $\chi^{(2)} (\varrho )$ denote two copies 
  of
 $\chi (\varrho )   $
%$\chi (\rho )\big| _{\rho=\varrho}  $ 
independent given $\varrho$, i.e.\ such that $\mathbb{E}\big[f(\chi
^{(1)} (\varrho ))g(\chi^{(2)} (\varrho ))\big|\varrho\big] = \mathbb{E}\big[f(\chi^{(1)} (\varrho ))\big|\varrho\big]\mathbb{E}\big[g(\chi^{(2)} (\varrho ))\big|\varrho\big]$ holds
for any Borel bounded functions $f$ and $g$.
The twin Monte Carlo validation procedure 
for a  predictor  $\Phi(\varrho)$ of $\Theta_0(\varrho)=\E(\chi(\varrho)|\varrho)$ 
consists in estimating by Monte Carlo
%(for each realization of the latter):
% For any Borel function $\varphi:\mathbb{R}^p\times\mathbb{R}^q\rightarrow\mathbb{R}$ such that $\varphi(X_i, Y_i)$ is square integrable (e.g.~a neural net estimate of $\mathbb{E}[\chi_{i,n}|X_i, Y_i]$), 
% we have:
\beql{e:val1}
\mathbb{E}\Big[\Phi%\Sigma^\theta_0 
(\varrho )^2-\big(\chi^{(1)}(\varrho )+\chi^{(2)}(\varrho )\big)\Phi (\varrho )+\chi ^{(1)}(\varrho )\chi^{(2)}(\varrho )\Big]
=
\mathbb{E}\Big[
\Big(\Phi%\Sigma^\theta_0
 (\varrho )-
 \mathbb{E}\big(\chi(\varrho ) \big|  \varrho  \big) 
%\Big(\mathbb{E}[\chi(\rho )]\big| _{\rho=\varrho} \Big)]
\Big)^2 \Big],\eeql
as it follows from the tower rule by conditional independence: see
Algorithm \ref{alg:twin}. % page \pageref{alg:twin}.
\begin{algorithm}[b!] 
 \def\then{m}
\def\thenu{m}
\def\thisq{f}
\def\this{g}
\def\un{\mathds{1}}
\small
\LinesNumbered
\SetAlgoLined
\SetKwInOut{AlgName}{name}
\SetKwInOut{Input}{input}\SetKwInOut{Output}{output}
%\AlgName{TwinVal}
\Input{out-of-sample $\{(\rho_j,\chi_j^{(1)}, \chi_j^{(2)})\}_{j = 1}^\thenu$ with independent draws $\rho_j$ of $\varrho$
and $\chi_j^{(1)}, \chi_j^{(2)}$ independent copies of $\chi$ given $\varrho=\rho_j$, $norm$ (set to $1$ by default)}%, a confidence level $\alpha$, corresponding estimates $\thisq$ and $\this$ of $q$ and $s$, tolerance levels $\delta^{\rm var}$ and $\delta^{\rm es}$}
 
\Output{Estimation of the normalized root mean square relative error and its $95\%$-confidence upper bound for a predictor $\Phi(\varrho)$ of $\Pi_0(\varrho)=\E(\xi|\varrho)$}% and $\this$} 
%Compute $(\epsilon^{\rm var})^2 = \frac{1}{\thenu}\sum_{i=1}^\thenu \big( (1-\alpha)(1-\alpha- \un_{\chi^{(1)}_j>\thisq (\rho_j)}- \un_{\chi^{(2)}_j>\thisq (\rho_j)}) + \un_{\chi^{(1)}\wedge \chi^{(2)}_j>\thisq (\rho_j)\big)} $ 
Compute $twin\mbox{-}stat   = \frac{1}{m} \sum_{j=1}^m \left[(\Phi(\rho_j))^2 - (\chi^{(1)}_j + \chi^{(2)}_j)\Phi(\rho_j) + \chi^{(1)}_j \chi^{(2)}_j\right]$

Compute $twin\mbox{-}stdev = \sqrt{\frac{1}{m}\sum_{j=1}^m\left[(\Phi(\rho_j))^2 - (\chi^{(1)}_j + \chi^{(2)}_j) \Phi(\rho_j)+ \chi^{(1)}_j \chi^{(2)}_j - twin\mbox{-}stat  \right]^2}$

Compute $twin\mbox{-}up = \sqrt{twin\mbox{-}stat +\frac{2}{\sqrt{m}} twin\mbox{-}stdev   }$

%\If{$norm$ not provided}{
% $norm$ = 1%Compute $norm = \sqrt{\frac{1}{m}\sum_{j=1}^m \chi^{(1)}_j \chi^{(2)}_j}$
%}

\uIf{$twin\mbox{-}stat > 0$}{

Compute $twin\mbox{-}err = \sqrt{twin\mbox{-}stat  }/ norm$

}\Else{

Set $twin\mbox{-}err$ to N/A

%Output N/A and $twin\mbox{-}up / norm$

}
Output $twin\mbox{-}err$ and $twin\mbox{-}ub :=twin\mbox{-}up / norm$

%}

\caption{Twin Monte Carlo validation for a predictor $\Phi(\varrho)$ of $\Pi_0(\varrho)=\E(\xi|\varrho)$.} 
 % As the estimator of the square twin error provided by a Monte Carlo estimate of the left-hand side in \eqref{e:val1} is not guaranteed to be positive, the estimated square error $twin\mbox{-}stat$ can be negative if the true value is small, in which case the relative error is expressed as N/A.}
\label{alg:twin} 
\end{algorithm}
The estimate $twin\mbox{-}stat$ of the square error can be negative in this algorithm. Thus, the $95\%$ upper bound of square error, remaining positive most of the times, is calculated alongside the square error itself. %Both Gaussian confidence interval (in principle only valid for Gaussian setup) and bootstrap methods (more general but slower to conclude) yields almost identical result.
We emphasize that the ensuing twin errors measure the performance of the predictor, but not of the associated sensitivities, as a low error of the predictor induces no constraint on the error for the derivatives. 

\begin{table}[t!]
    \centering
    \resizebox{\textwidth}{!}{
    \begin{tabular}{|l|ccc|ccc|}
\hline
\multirow{2}{*}{$t$ (in yr)} & \multicolumn{3}{c|}{Baseline mode} & \multicolumn{3}{c|}{Risk mode} \\ \cline{2-7} 
 & \multicolumn{1}{c|}{Nested MC} & \multicolumn{1}{c|}{NN} & Linear & \multicolumn{1}{c|}{Nested MC} & \multicolumn{1}{c|}{NN} & Linear \\ \hline
$0.01$ & \multicolumn{1}{c|}{N/A(4.3\%)} & \multicolumn{1}{c|}{N/A(4.3\%)} & N/A(4.5\%) & \multicolumn{1}{c|}{N/A(4.3\%)} & \multicolumn{1}{c|}{2.1\%(5.0\%)} & 3.0\%(5.5\%) \\ \hline
$0.1$ & \multicolumn{1}{c|}{6.6\%(8.0\%)} & \multicolumn{1}{c|}{5.7\%(7.3\%)} & 6.8\%(8.2\%) & \multicolumn{1}{c|}{6.6\%(8.0\%)} & \multicolumn{1}{c|}{9.8\%(10.8\%)} & 9.8\%(10.8\%) \\ \hline
1 & \multicolumn{1}{c|}{9.2\%(10.3\%)} & \multicolumn{1}{c|}{9.7\%(10.8\%)} & 10.4\%(11.4\%) & \multicolumn{1}{c|}{9.4\%(10.5\%)} & \multicolumn{1}{c|}{22.1\%(22.6\%)} & 22.3\%(22.9\%) \\ \hline
$3.5$ & \multicolumn{1}{c|}{8.8\%(10.1\%)} & \multicolumn{1}{c|}{12.0\%(13.5\%)} & 15.7\%(17.3\%) & \multicolumn{1}{c|}{9.7\%(11.5\%)} & \multicolumn{1}{c|}{26.0\%(27.0\%)} & 27.0\%(28.2\%) \\ \hline
\end{tabular}}
    \caption{The mean and ($95\%$ upper bound) for the %out-of-sample 
    relative twin error normalized by $\CVA_0(\rho_0)$  (cf.\ Algorithm \ref{alg:twin}) of nested Monte Carlo $\CVA_t$, neural network $\CVA_t$, and linear regression $\CVA_t$, for $t=0.01, 0.1, 1 $, and $3.5\approx T/3 $ years (where the CVA of a portfolio of swaps is deemed the most volatile), in the baseline and risk modes.}
    \label{tab:cva_error}
\end{table}    
Table \ref{tab:cva_error} % page \pageref{tab:cva_error}
shows the twin scores of neural network and linear regression versus nested Monte Carlo predictors of CVA$_t(X_t,\varrho_t)$ in the baseline and risk modes. In both  modes, all three methods have comparable performance for small $t$ ($0.01$ and $0.1$ years). The linear learning model for $\CVA^\theta_t(x,\rho)$ does not require sophisticated training like the neural network but only linear algebra, it converges equally well for $t=0.01$ and $0.1$, but falls short at one year, where the nonlinearity of CVA$_t(x,\rho)$ becomes significant: 
for $t=1$ and $3.5$ years, the neural network outperforms the linear regression significantly.
As illustrated by the left panel of Figure 1 in \citep[Section 2]{abbas2021hierarchical}, this nonlinearity and the necessity of neural network would become more stringent with option portfolios, possibly for lower $t$ already. In the baseline mode, the learned $\CVA_t(X_t,\varrho_t)$ can occasionally be more accurate than its nested Monte Carlo counterpart, which is also much slower: for a given $t$, the nested Monte Carlo CVA  (with  $m=2^{17}$ outer paths and 1024 inner paths throughout the paper) takes approximately 40 minutes, while simulating the data and training a neural net CVA with 2 hidden layers and softplus activation (resp.\ regressing a linear CVA) takes roughly 30 (resp.\ 25) seconds.
%\citep[Table 3, Section 5.1]{albanese2021xva}. 
In the risk mode, the neural network is less accurate than nested Monte Carlo: the input dimension of the neural network becomes 
35 for $(x,y)$ plus 63 for $\epsilon$ in the risk mode versus 35 simply in the baseline mode, hence training $\CVA^\theta_t(x,\rho)$ becomes harder (or would require more data) and the resulting predictor becomes less accurate than nested Monte Carlo. For $t=3.5$ the neural network prediction is bad, but not as bad as the linear prediction. A better neural network predictor might be achievable by fine-tuning the SGD or enriching the simulated dataset and/or the architecture of the network. Since we only need the risk mode for $t\le 1 
$ in the dowstream tasks of Sections \ref{ss:riskdyn}-\ref{s:risk},
% (see the introduction to Part \ref{p:II}),  
we did not venture in these directions.

\FloatBarrier

\section{Run-off CVA Risk\label{ss:riskdyn}}

An economic (or ``internal'') view gained from simulating the movements of model or/and market risk factors and obtaining risk measures of CVA fluctuations is an important dimension of the CVA capital regulatory requirements of a bank, in the context of its supervisory review and evaluation process (SREP): 
quoting \url{https://www.bankingsupervision.europa.eu/legalframework/publiccons/html/icaap_ilaap_faq.en.html} (last accessed June 6 2024), ``the risks the institution has identified and quantified  will play an enhanced role in, for example, the determination of additional own funds requirements on a risk-by-risk basis.''
% In \citep[article 50.15(1), top of page 11]{BaselIICIVAnew},
% $t=1$ is mentioned as the reference risk horizon that should be used in CVA risk calculations, but such recommendation on the value of $t$ is no longer present in the updated version \citep{BaselIICIVAbrandnew}. 
% A CVA risk horizon as large as one  year can indeed be relevant for credit derivatives portfolios.
% In our experiments, in order to assess the impact on the results of the CVA
% nonlinearity, which is more sizeable with a longer risk horizon,
% we consider three different CVA risk horizons: $t=0.01,$ $0.1$ and 1 yr.

The focus of this section is on 
\beql{e:d_cva_risk} \delta\CVA^{\theta}_t +\LGD_t\sp \mbox{where }  \delta\CVA^{\theta}_t = \CVA^{\theta}_t (X_t,\varrho_t)  - \CVA_0(\rho_0)  \eeql
(cf.\  \eqref{eq:LGD}-\eqref{eq:CVAint}), assessed in the risk mode, where $\varrho_t = (Y_t(y_0,\varepsilon), \varepsilon) $ and $\varepsilon$ follows $\mathcal{N}\Big(\epsilon_0, \Id \left(t(1\%)^2\epsilon_0 \odot \epsilon_0 \right) \Big)$.
The random variable ($\delta\CVA^{\theta}_t +\LGD_t$) reflects a dynamic but also run-off view on CVA and counterparty default risk altogether, as opposed to the stationary run-on CVA risk view of Section \ref{s:risk}. We assess CVA risk, counterparty default risk, and both risks combined, on the basis of value-at-risks (VaR) and expected shortfalls (ES) of $\delta\CVA^{\theta}_t$ and LGD$_t$ in the risk mode, reported for $t=0.01, 0.1, 1$ yr and various quantile levels $\alpha= 95, 97.5$ or $99\%$ in Table \ref{t:risk_measure}. 
The middle-point of $97.5\%$  for the ES fits a nowadays reference $99.9\%$ value-at-risk reference level, via the conventional mapping between a $97.5\%$ ES and a  $99\%$ VaR in a Gaussian setup, also considering that our baseline parametrization reflects market conditions at a 90\% level of stress (cf.\ the high CDS spreads visible in the bottom left panel of Figure \ref{fig:zc_cds_sensis}).
The results of Table \ref{t:risk_measure} page \pageref{t:risk_measure} emphasize that CVA and counterparty default risks assessed on a run-off basis are primarily driven by client defaults, especially at higher quantile levels. As visible in the right plots, for $t=0.01$ and  $0.1$, there are few client defaults and the right tail of the distribution of $\delta \CVA^\theta_t + \LGD_t$ is dominated by the term $\delta \CVA^\theta_t$.  For $t=1$, instead, the $\LGD_t$ term takes the lead, significantly shifting the right tail of the distribution upward. 
 \begin{table}[t!]
\begin{minipage}{0.5\textwidth}
\resizebox{\textwidth}{!}{
 \begin{tabular}{|l|r|r|r|}
\hline
$t=0.01$ & \multicolumn{1}{c|}{$\delta \CVA^\theta_t$} & \multicolumn{1}{c|}{$\LGD_t$} & \multicolumn{1}{c|}{\begin{tabular}[c]{@{}c@{}}$\delta \CVA^\theta_t$\\ $+ \LGD_t$\end{tabular}} \\ \hline
Expectation & -9 & 0.28 & -8 \\ \hline
VaR 95\% & 268 & 0 & 268 \\ %\hline
VaR 97.5\% & 323 & 0 & 323 \\ %\hline
VaR 99\% & 388 & 0 & 389 \\ \hline
ES 95\% & 344 & 0.28 & 345 \\ %\hline
ES 97.5\% & 395 & 0.28 & 397 \\ %\hline
ES 99\% & 460 & 0.28 & 464 \\ \hline
\end{tabular}}
\end{minipage}%
\begin{minipage}{0.45\textwidth}
\centering
    \includegraphics[width = \textwidth]{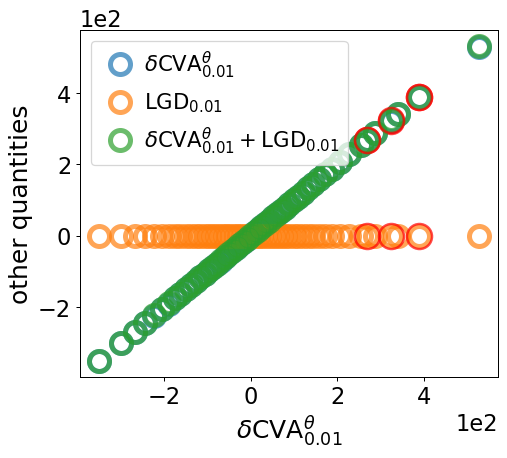}
\end{minipage}
\begin{minipage}{0.5\textwidth}
\resizebox{\textwidth}{!}{
\begin{tabular}{|l|r|r|r|}
\hline
$t=0.1$ & \multicolumn{1}{c|}{$\delta \CVA^\theta_t$} & \multicolumn{1}{c|}{$\LGD_t$} & \multicolumn{1}{c|}{\begin{tabular}[c]{@{}c@{}}$\delta \CVA^\theta_t$\\ $+ \LGD_t$\end{tabular}} \\ \hline
Expectation & 56 & 7 & 63 \\ \hline
VaR 95\% & 939 & 0 & 953 \\ %\hline
VaR 97.5\% & 1,124 & 0 & 1,145 \\ %\hline
VaR 99\% & 1,347 & 0 & 1,389 \\ \hline
ES 95\% & 1,192 & 7 & 1,243 \\ %\hline
ES 97.5\% & 1,361 & 7 & 1,445 \\ %\hline
ES 99\% & 1,571 & 7 & 1,740 \\ \hline
\end{tabular}}
\end{minipage}%
\begin{minipage}{0.45\textwidth}
\centering
    \includegraphics[width = \textwidth]{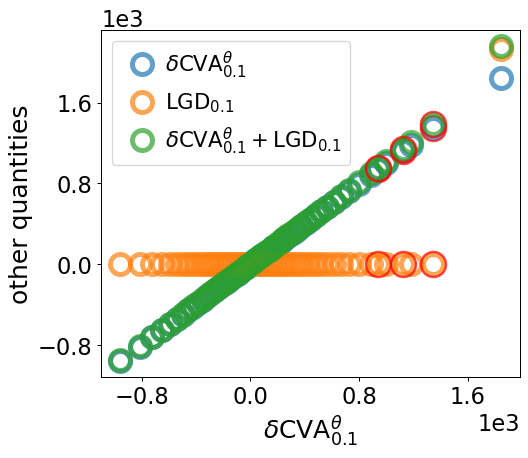}
\end{minipage}

\begin{minipage}{0.5\textwidth}
\resizebox{\textwidth}{!}{
\begin{tabular}{|l|r|r|r|}
\hline
$t=1$ & \multicolumn{1}{c|}{$\delta \CVA^\theta_t$} & \multicolumn{1}{c|}{$\LGD_t$} & \multicolumn{1}{c|}{\begin{tabular}[c]{@{}c@{}}$\delta \CVA^\theta_t$\\ $+ \LGD_t$\end{tabular}} \\ \hline
Expectation & 75 & 502 & 578 \\ \hline
VaR 95\% & 2,942 & 3,621 & 4,597 \\ %\hline
VaR 97.5\% & 3,746 & 7,309 & 7,190 \\ %\hline
VaR 99\% & 4,796 & 11,997 & 11,757 \\ \hline
ES 95\% & 4,122 & 8,846 & 8,980 \\ %\hline
ES 97.5\% & 4,953 & 12,383 & 12,297 \\ %\hline
ES 99\% & 6,102 & 17,004 & 17,048 \\ \hline
\end{tabular}
}
\end{minipage}%
\begin{minipage}{0.45\textwidth}
\centering
    \includegraphics[width = \textwidth]{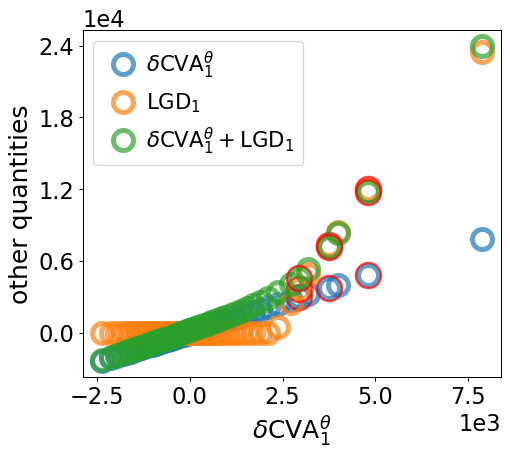}
\end{minipage}
\caption{$\delta\CVA^{\theta}_t,$ LGD$_t$ and their sum ($\CVA_0(\rho_0)=5,027$). (\textit{Left}) Risk measures. (\textit{Right}) qq plots
%(of the whole distribution) 
against $\delta \CVA^\theta_t$ used as a benchmark (corresponding to the diagonal).
The red circles correspond to the VaRs  95\%, 97.5\%
 and 99\%. 
}\label{t:risk_measure}

\end{table}

%\FloatBarrier
\subsection{Run-off CVA Hedging\label{ss:hoff}}
\def\Theta{c}
Let (in vector form) $\CF$  represent the cumulative cash flows process of the market instruments of Section \ref{s:toymo}, with price process $Z_t=\E (\CF_T -\CF_t | X_t,\varrho_t), t\in  0\,..\, nh=T$, and
let \beql{e:Zt}\delta Z_t = Z _t(X_t,\varrho_t)- z_0\eeql  denote the difference between the market prices $Z _t(X_t,\varrho_t)$ 
and their time-0 baseline price $z_0$.
By loss $L^{\theta}_{\theh} $, we mean the following hedged loss(-and-profit) of the CVA desk over the risk horizon $t$:
\beql{e:losstheZ}  
 L^{\theta}_{\theh} = \delta  \CVA^{\theta}_t + \LGD_{\theh}   - (\delta Z_t + \CF_t)^\top\Delta 
%\Delta  (\delta Z_t + \CF_t) 
 -\Theta ,
\eeql      
where the hedging ratio $\Delta \in \mathbb{R}^{ q   } $ is treated as a free parameter, while $\Theta$ is deduced from $\Delta$ through the constraint that ${\mathbb{E}} L^{\theta}_{\theh}=0 $  (or $\widehat{\mathbb{E}} L^{\theta}_{\theh}=0 $ in our numerics).
The constant $c$,  which is equal to 0 (modulo the numerical noise) in the baseline mode where $\CVA  + \LGD   $ and $Z  + \CF$  are both martingales, can be interpreted in terms of a hedging valuation adjustment (HVA) in the spirit of 
\citet*{AlbaneseCrepeyBenezet22}, i.e.\ a provision for model risk. \citet{AlbaneseCrepeyBenezet22} develop how, such a provision having been set apart in a first stage, the loss $L^{\theta}_{\theh}$, thus centered via the ``HVA trend'' $c$ 
(i.e. $c=-\E\delta {\rm HVA}_t={\rm HVA}_0-\E{\rm HVA}_t$),
deserves an economic capital, which we quantify below as an expected shortfall of $L^{\theta}_{\theh}$. 

Bump sensitivities $\Delta=\Delta^{bump}$ can be used in \eqref{e:losstheZ} but they are expected to be inappropriate for dealing with client defaults.
As an alternative approach yielding hedging ratios, HVA trend and economic capital at the same time, one can use
\begin{equation}\label{eq:ESMinoff}
     \Delta^{ec} = \argmin\limits_{ \Delta  \in \mathbb{R}^q } \ES\left( L^{\theta}_{t} \right).
\end{equation} 
Following \citeN{rockafellar2000optimization}, \eqref{eq:ESMinoff}  can be reformulated as the following convex optimization problem:
\begin{equation}\label{bayeq:ESMinRockafellaroff}
    \left( \Delta^{ec}  , k^{ec}\right) = \argmin\limits_{ \Delta   \in \mathbb{R}^q , \strike\in \mathbb{R}}  \strike + (1 - \alpha)^{-1} \mathbb{E} \left[ \left( %\tilde 
    L^{\theta}_{t}
    % \b{-\mathbb{E}L^{\theta}_{\theh}}
     - \strike
       \right)^{+} \right] ,
\end{equation}   
where $k^{ec}$ is then the value-at-risk (quantile of level $\alpha$) of the corresponding $ L^{\theta}_{t}$.
Note that one could also easily account for transaction costs in this setup, which is nothing but a deep hedging approach over one time step for computing the HVA trend $c$ and the hedging policy $\Delta^{ec}$ compressing the ensuing economic capital and its 
%KVA 
cost.
%\citep{Buehler2017StatisticalH,BuhlerGononTeichmannWood18,AlbaneseCrepey16}. 
Note that
the constraint $\mathbb{E} L^{\theta}_{\theh}=0$, motivated financially after \eqref{e:losstheZ}, is also necessary  numerically to stabilize the training of the EC sensitivities.

As a possible (simpler) variation on
the \textbf{economic capital (EC) run-off sensitivities} \eqref{eq:ESMinoff}-\eqref{bayeq:ESMinRockafellaroff}, we also consider the following \textbf{PnL explain 
(\PLE) run-off sensitivities}: 
\begin{equation}\label{bayeq:LSoff}
     \Delta^{ple}  
    = \argmin\limits_{ \Delta \in \mathbb{R}^q
    }  \mathbb{E}  [  (%\tilde 
    L^{\theta}_{t} %\b{-c}
    )^2] . 
\end{equation} 
Once $\CVA^{\theta}_t (X_t,\varrho_t) $ learned from the cash flows $\xi_{t,T}(\varrho_t)$ the way described in Algorithm \ref{alg:cva_learning} / risk mode, 
the EC sensitivities \eqref{eq:ESMinoff} are computed by stochastic gradient descent based on an empirical version of \eqref{bayeq:ESMinRockafellaroff}; the \PLE sensitivities are computed 
by performing a linear regression
%\footnote{implemented using a truncated singular value decomposition (SVD) approach, see~e.g.~\citeN[Theorems 2.5.2 and 5.5.1]{golub2013matrix}.} 
corresponding to an empirical version of \eqref{bayeq:LSoff}:
cf.\ lines 9--13 of 
%the analogous run-on 
Algorithm \ref{alg:reg_sensi} page \pageref{alg:reg_sensi} (which corresponds to the
CVA run-on setup of Section \ref{s:risk}).

By
unexplained \PnL UPL (resp.\ economic capital EC),  
we mean the standard error (resp.\ expected shortfall) of  $L^{\theta}_{t}$.
As performance metrics, we consider a backtesting, out-of-sample UPL 
(resp.\ EC with $\alpha = 95\%$) for $\Delta=\bo $, divided by 
 UPL (resp.\ EC with $\alpha = 95\%$) for each considered  set of sensitivities:
% As performance metrics, we consider a backtesting (i.e.\ out-of-sample) 
% %standard error of the 
% unexplained \PnL (resp.\ economic capital for $\alpha = 95\%$) 
%  of $\delta\CVA^{\theta}_{t}$, i.e.\ of $L^{\theta}_{t}$ for $\Delta=\bo  $, divided by the 
%  %standard error of the 
%  unexplained \PnL  (resp.\ economic capital for $\alpha = 95\%$) of  $L^{\theta}_{t}$ as per \eqref{e:losstheZ}, for each considered  set of sensitivities:
 the higher the corresponding ``compression ratios'', the better the corresponding sensitivities. 
%, featuring different levels of nonlinearity of the CVA.
For each simulation run below (here and in Section \ref{s:risk}), we use $m = 2^{17}$ paths to estimate the \PLE and EC sensitivities %by Algorithm \ref{alg:reg_sensi} 
 and we generate other $m = 2^{17}$ paths for our backtest. 
 % All the numerical results below are thus out-of-sample, \b{in line with our backtesting purposes.}

\begin{figure}[t!]
    \centering
    \includegraphics[width = \textwidth, height = 499pt ]{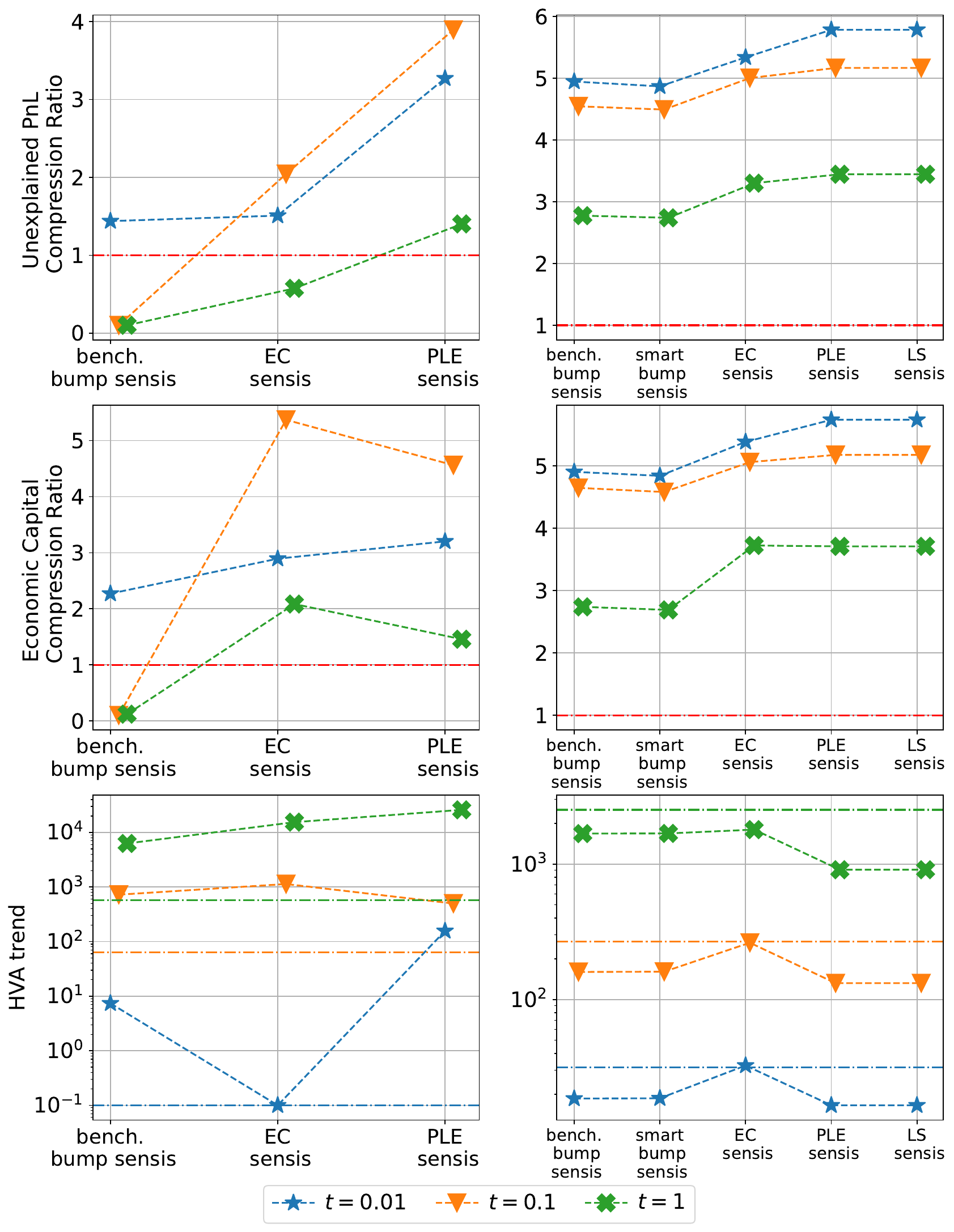}
    \caption{Compression ratios of UPL (\textit{top}), EC (\textit{middle}), and HVA trends $c$ (\textit{bottom}), for various $\delta\CVA^{\theta}_t +\LGD_t$ hedging approaches in Section \ref{ss:hoff} (\textit{left}), and $\delta\CVA^{\theta}_{(t)}$ hedging approaches in Section \ref{s:cpes} (\textit{right}). Horizontal dash-dot lines correspond to the unhedged case with $\Delta = \bo$ in \eqref{e:losstheZ} and \eqref{e:lossZ}.}
    \label{fig:risk_hedge}
\end{figure}

The left panels in Figure \ref{fig:risk_hedge} compare the hedging performance of benchmark bump versus EC or \PLE sensitivities for $t=0.01, 0.1,$ and $1$ yr.
The unhedged case corresponds to the red horizontal dash-dot lines. Consistent with their definitions, the \PLE sensitivities always (even though we are out-of-sample) yield the highest unexplained \PnL compression ratios, while the EC sensitivities, except for $t=0.01$, yield the highest EC compression ratio. As expected, bump sensitivities provide very poor hedging performance (we omitted fast bump sensitivities, which provide results similar to
%as there is no reason why they should outperform 
the benchmark ones). At the risk horizon $t=0.01$, client defaults are rare (only occurring on $0.017\%$ of the scenarios) and a bump sensitivities hedge reduces risk, but much less so than the EC or \PLE sensitivities hedges. 
For $t=0.1$ and 1 yr, hedging by bump sensitivities is even counterproductive, worsening both unexplained \PnL and economic capital compared to the 
unhedged case; EC and \PLE sensitivities hedges achieve significant 
unexplained \PnL and economic capital
compression, but this comes along with very high HVA trends $c$.
%, even higher in the $t= 1$ yr case than  the corresponding unhedged EC levels (retrievable from the $\delta \CVA^\theta_t+ \LGD_t$ columns of Table \ref{t:risk_measure} as the differences between the ES 95\% and Expectation lines). 

The expected conclusion of this part is that for properly hedging CVA in the run-off mode, one should first
replicate the impact of the defaults with appropriate CDS positions (or decide to warehouse default risk, especially if CDSs are not liquidly available),
%(provided these would be liquidly available), 
rather than trying to hedge ``on average'', which only makes sense for CVA assessed in the run-on mode. The latter corresponds to the right panels in Figure \ref{fig:risk_hedge}, to be commented upon in  Section \ref{s:cpes}.
%(or perhaps at a very short hedging horizon $t$).}
% \b{The results of this section are in line with the ones obtained in a different context of SIMM computations
% %and focused on a quadratic hedging criterion,  
% by \citeN{albanese2017var}: bump sensitivities may be
% too local for dealing with CVA risk, at least if meant in a run-off mode mainly driven by granular client default events. 
% % Optimized PnL explain sensitivities are better for this purpose, but they can only be derived by learning the CVA in the first place.
% %  This learned CVA itself is a better input to CVA risk computations than any linear or even linear quadratic proxy.  

% Under an alternative CVA deep hedging approach, one would look simultaneously for the value of the CVA and the corresponding dynamic hedging policy jointly minimizing some risk measure of the CVA trader's hedged loss at the final maturity of the portfolio. The size of the corresponding training problem would be of a completely different order of magnitude than the ones involved in this work and such a CVA deep hedging approach approach would not scale at the level of realistic banking portfolios.
%  Moreover the corresponding CVA is no longer the expectation (under the same ambient measure) of the CVA originating cash flows, i.e.\ it departs from
%  the specification
%   \eqref{eq:CVAint}-\eqref{e:lecva}. For these two reasons we refrain from venturing in this direction.

\FloatBarrier

\section{Run-on CVA Risk}\label{s:risk}
 %CVA Risk Assessed and Hedged on a Run-On Basis

With $\cdot_{(t)}$ in $\varepsilon
%_0
_{(t)}$  referring to the dependence of the variance $\sigma^2$ of $\varepsilon
-\epsilon_0$ (in the notation of Algorithm \ref{alg:cva_learning}) in $t$ in what follows, namely $\Var (\varepsilon
%_0
_{(t)})=t\times 0.01\%$, denoting by $Y_t(y_0,\varepsilon_{(t)})$ the $Y$ process at time $t$ starting from $y_0$ at 0 and for model parameters set to $\varepsilon_{(t)}$, and by $\varrho_{(t)}=\big(Y_t(y_0,\varepsilon _{(t)}),\varepsilon _{(t)}\big)$,
let
(cf.\ \eqref{e:Zt}) 
\beql{e:pl_cva_sensi}
&
  \delta \varrho_{(t)} =( Y _t(y_0,\varepsilon  _{(t)})- y_0 , \varepsilon  _{(t)}- \epsilon_0)
=\varrho_{(t)}-\rho_0\sp \delta Z_{\ot}=Z _0 (\varrho_{(t)})- z_0 , \\ 
%\delta \varepsilon = \varepsilon_{(t)} - \epsilon_0 \mbox{, and}\\ 
&
\delta  \CVA_{\ot}%^{\theta} 
= \CVA%^{\theta}
_0 (\varrho_{(t)}) -
\CVA_0(\rho_0).
\eeql 
 The fact that we consider the time-0 $\CVA_0 (\varrho_{(t)})$ (see after \eqref{eq:CVAint}) and likewise $Z _0 (\varrho_{(t)})$
 %(skipping $X_0 \equiv 0 $ from the conditioning)
 %in the sensis mode 
here, as opposed to
 $\CVA_t  (X_t,\varrho_t)$ and  $Z _t (X_t,\varrho_t)$ 
 in Section \ref{ss:riskdyn},
 is in line with an assessment of risk on a run-on portfolio and customers basis and with a siloing of CVA vs.\ counterparty default risk, which have both become standard in regulation and market practice. 
Various predictors of $\delta \CVA_{\ot}$ can be learned directly from the simulated model parameters $\varrho_{\ot}$ and 
cash flows 
$\xi_{0,T}(\varrho_{(t)})-\xi_{0,T}(\rho_0) $ 
%in the sensis mode 
(as opposed to and better than learning $\delta \CVA_{\ot}$ via $ \CVA_{0} (\varrho_{(t)})$, which would involve more variance): nested Monte Carlo estimator, neural net regressor $\delta \CVA^\theta_{\ot}$, linear(-diagonal quadratic) regressors against $\delta \varrho_{(t)}$ or $\delta Z_{(t)}$ referred to as LS (for ``least squares'') below. 
 The neural network used for training
$\delta\CVA^\theta_{\ot}$
 based on simulated data $(\varrho_{(t)} -\rho_0,   \xi_{0,T}(\varrho_{(t)};\omega) - \xi_{0,T}(\rho_0;\omega))$ as per line 8 of Algorithm \ref{alg:reg_sensi}
 has one hidden layer with two hundred hidden units and softplus activation functions.
 
Table \ref{t:sensi} displays some twin upper bounds and risk measures of $\delta \CVA_{\ot}$ computed with 
%and the twin error of various $\delta \CVA_{\ot}$
these different approximations, as well as with linear(- diagonal quadratic) Taylor expansions in  $\delta \varrho_{\ot} $ or $\delta Z_{\ot}$ with coefficients estimated as benchmark or \smart bump sensitivities. 

In terms of the twin upper bounds, the nested CVA has the best accuracy, but (for a given risk horizon $t$) it takes about 2 hours, versus  about 
one minute of simulation time for generating the labels,
plus 30 seconds for training by neural networks and 2-3 seconds for LS regression. The neural network excels at large $t$, where the non-linearity becomes significant, while being outperformed by the linear methods at small $t$, where $\delta \CVA_{\ot}$ is approximately linear. With diagonal gamma ($\Gamma$) elements taken into account, the performance of the LS regressor improves at large $t$.
The linear quadratic Taylor expansions show relatively good twin upper bounds for small $t$, but worsen for large $t$.

Regarding now the risk measures, the nested $\delta \CVA_{\ot}$ and the neural network  $\delta \CVA^\theta_{\ot}$  
provide more conservative VaR and ES estimates than any linear(-quadratic) proxy in all cases. Surprisingly, even though $\CVA%^{\theta}
_0 (\varrho_{(t)})$ in $\delta \CVA_{\ot}$ is a function of $\varrho_{\ot}$, for $t =1$ the linear(-quadratic) proxies in $\delta Z_{\ot}$ outperform those in $\varrho_{\ot}$,
in terms both of twin error and of consistency of the ensuing risk measures with those provided by the 
nonparametric
(neural net and nested) references. 
Also note that, when compared with the nonparametric approaches again, the \smart bump sensitivities proxy in $\delta Z_{\ot}$ yields results almost as good as the much slower benchmark bump sensitivities proxy in $\delta Z_{\ot}$ (see Tables \ref{tab:timesensis} and \ref{tab:time} pages \pageref{tab:timesensis} and \pageref{tab:time}).
\begin{table}[t]
\resizebox{\textwidth}{!}{
\begin{tabular}{|c|l|rr|rrr|rrr|}
\hline
\multirow{2}{*}{$t$} & \multicolumn{1}{c|}{\multirow{2}{*}{\begin{tabular}[c]{@{}c@{}}risk\\ measure\end{tabular}}} & \multicolumn{2}{c|}{nonparametric} & \multicolumn{3}{c|}{linear quadratic in $\delta\varrho_{\ot}$} & \multicolumn{3}{c|}{linear quadratic in $\delta Z_{\ot}$ } \\ \cline{3-10} 
 & \multicolumn{1}{c|}{} & \multicolumn{1}{c|}{\begin{tabular}[c]{@{}c@{}}nested\\ $\delta \CVA_{\ot}$\end{tabular}} & \multicolumn{1}{c|}{$\delta \CVA^\theta_{\ot}$} & \multicolumn{1}{c|}{\begin{tabular}[c]{@{}c@{}}bench.\\ bump\\ sensis w/ $\Gamma$\end{tabular}} & \multicolumn{1}{c|}{\begin{tabular}[c]{@{}c@{}}\smart\\ bump\\ sensis\end{tabular}} & \multicolumn{1}{c|}{\begin{tabular}[c]{@{}c@{}}LS sensis\\ w/ $\Gamma$\end{tabular}} & \multicolumn{1}{c|}{\begin{tabular}[c]{@{}c@{}}bench.\\ bump\\ sensis\end{tabular}} & \multicolumn{1}{c|}{\begin{tabular}[c]{@{}c@{}}\smart\\ bump\\ sensis\end{tabular}} & \multicolumn{1}{c|}{\begin{tabular}[c]{@{}c@{}}LS sensis\\ w/ $\Gamma$\end{tabular}} \\ \hline
\multirow{7}{*}{0.01} & twin-ub & \multicolumn{1}{r|}{\textbf{11}} & 29 & \multicolumn{1}{r|}{\textbf{13}} & \multicolumn{1}{r|}{\textbf{15}} & 18 & \multicolumn{1}{r|}{23} & \multicolumn{1}{r|}{23} & 18 \\ \cline{2-10} 
 & VaR 95\% & \multicolumn{1}{r|}{\textbf{364}} & \textbf{315} & \multicolumn{1}{r|}{312} & \multicolumn{1}{r|}{306} & 302 & \multicolumn{1}{r|}{\textbf{315}} & \multicolumn{1}{r|}{312} & 308 \\
 & VaR 97.5\% & \multicolumn{1}{r|}{\textbf{431}} & \textbf{369} & \multicolumn{1}{r|}{366} & \multicolumn{1}{r|}{358} & 354 & \multicolumn{1}{r|}{\textbf{370}} & \multicolumn{1}{r|}{367} & 362 \\
 & VaR 99\% & \multicolumn{1}{r|}{\textbf{510}} & \textbf{431} & \multicolumn{1}{r|}{429} & \multicolumn{1}{r|}{418} & 415 & \multicolumn{1}{r|}{\textbf{435}} & \multicolumn{1}{r|}{431} & 424 \\ \cline{2-10} 
 & ES 95\% & \multicolumn{1}{r|}{\textbf{454}} & \textbf{387} & \multicolumn{1}{r|}{384} & \multicolumn{1}{r|}{375} & 372 & \multicolumn{1}{r|}{\textbf{389}} & \multicolumn{1}{r|}{385} & 380 \\
 & ES 97.5\% & \multicolumn{1}{r|}{\textbf{514}} & \textbf{434} & \multicolumn{1}{r|}{432} & \multicolumn{1}{r|}{421} & 417 & \multicolumn{1}{r|}{\textbf{437}} & \multicolumn{1}{r|}{433} & 427 \\
 & ES 99\% & \multicolumn{1}{r|}{\textbf{585}} & \textbf{491} & \multicolumn{1}{r|}{489} & \multicolumn{1}{r|}{476} & 473 & \multicolumn{1}{r|}{\textbf{495}} & \multicolumn{1}{r|}{490} & 484 \\ \hline
\multirow{7}{*}{0.1} & twin-ub & \multicolumn{1}{r|}{\textbf{59}} & 103 & \multicolumn{1}{r|}{110} & \multicolumn{1}{r|}{111} & 109 & \multicolumn{1}{r|}{\textbf{101}} & \multicolumn{1}{r|}{104} & \textbf{94} \\ \cline{2-10} 
 & VaR 95\% & \multicolumn{1}{r|}{\textbf{1,223}} & \textbf{1,189} & \multicolumn{1}{r|}{\textbf{1,154}} & \multicolumn{1}{r|}{1,101} & 1,148 & \multicolumn{1}{r|}{1,147} & \multicolumn{1}{r|}{1,138} & 1,151 \\
 & VaR 97.5\% & \multicolumn{1}{r|}{\textbf{1,431}} & \textbf{1,383} & \multicolumn{1}{r|}{\textbf{1,340}} & \multicolumn{1}{r|}{1,266} & 1,327 & \multicolumn{1}{r|}{1,330} & \multicolumn{1}{r|}{1,318} & 1,332 \\
 & VaR 99\% & \multicolumn{1}{r|}{\textbf{1,686}} & \textbf{1,618} & \multicolumn{1}{r|}{\textbf{1,562}} & \multicolumn{1}{r|}{1,463} & 1,539 & \multicolumn{1}{r|}{1,554} & \multicolumn{1}{r|}{1,542} & 1,556 \\ \cline{2-10} 
 & ES 95\% & \multicolumn{1}{r|}{\textbf{1,504}} & \textbf{1,449} & \multicolumn{1}{r|}{\textbf{1,403}} & \multicolumn{1}{r|}{1,321} & 1,386 & \multicolumn{1}{r|}{1,396} & \multicolumn{1}{r|}{1,383} & 1,396 \\
 & ES 97.5\% & \multicolumn{1}{r|}{\textbf{1,693}} & \textbf{1,622} & \multicolumn{1}{r|}{\textbf{1,570}} & \multicolumn{1}{r|}{1,467} & 1,544 & \multicolumn{1}{r|}{1,559} & \multicolumn{1}{r|}{1,545} & 1,561 \\
 & ES 99\% & \multicolumn{1}{r|}{\textbf{1,923}} & \textbf{1,830} & \multicolumn{1}{r|}{\textbf{1,769}} & \multicolumn{1}{r|}{1,638} & 1,731 & \multicolumn{1}{r|}{1,757} & \multicolumn{1}{r|}{1,740} & 1,757 \\ \hline
\multirow{7}{*}{1} & twin-ub & \multicolumn{1}{r|}{\textbf{325}} & \textbf{693} & \multicolumn{1}{r|}{1,244} & \multicolumn{1}{r|}{1,307} & 1,113 & \multicolumn{1}{r|}{932} & \multicolumn{1}{r|}{943} & \textbf{743} \\ \cline{2-10} 
 & VaR 95\% & \multicolumn{1}{r|}{\textbf{7,097}} & \textbf{6,992} & \multicolumn{1}{r|}{6,365} & \multicolumn{1}{r|}{5,300} & 6,440 & \multicolumn{1}{r|}{5,867} & \multicolumn{1}{r|}{5,831} & \textbf{6,676} \\
 & VaR 97.5\% & \multicolumn{1}{r|}{\textbf{8,433}} & \textbf{8,199} & \multicolumn{1}{r|}{7,451} & \multicolumn{1}{r|}{5,897} & 7,432 & \multicolumn{1}{r|}{6,737} & \multicolumn{1}{r|}{6,686} & \textbf{7,838} \\
 & VaR 99\% & \multicolumn{1}{r|}{\textbf{10,333}} & \textbf{9,887} & \multicolumn{1}{r|}{8,991} & \multicolumn{1}{r|}{6,646} & 8,812 & \multicolumn{1}{r|}{7,914} & \multicolumn{1}{r|}{7,846} & \textbf{9,493} \\ \cline{2-10} 
 & ES 95\% & \multicolumn{1}{r|}{\textbf{9,163}} & \textbf{8,805} & \multicolumn{1}{r|}{8,077} & \multicolumn{1}{r|}{6,142} & 7,956 & \multicolumn{1}{r|}{7,162} & \multicolumn{1}{r|}{7,097} & \textbf{8,492} \\
 & ES 97.5\% & \multicolumn{1}{r|}{\textbf{10,654}} & \textbf{10,090} & \multicolumn{1}{r|}{9,309} & \multicolumn{1}{r|}{6,715} & 9,032 & \multicolumn{1}{r|}{8,075} & \multicolumn{1}{r|}{7,988} & \textbf{9,792} \\
 & ES 99\% & \multicolumn{1}{r|}{\textbf{12,781}} & \textbf{11,869} & \multicolumn{1}{r|}{11,161} & \multicolumn{1}{r|}{7,456} & 10,598 & \multicolumn{1}{r|}{9,339} & \multicolumn{1}{r|}{9,216} & \textbf{11,677} \\ \hline
\end{tabular}
}
\caption{Risk measures of $\delta \CVA_{(t)}$ computed by Monte Carlo using $\delta \CVA_{(t)}$ simulated by various predictors. The three lowest (i.e.\ best) twin errors (without normalization, cf.\ Algorithm \ref{alg:twin} page \pageref{alg:twin}) and the three highest (i.e. most conservative) risk estimates on each row  
are emphasized in bold.}  \label{t:sensi}
\end{table}

\FloatBarrier

\subsection{Run-on CVA Hedging}\label{s:cpes}
% the Run-on CVA
Let 

\beql{e:lossZ} 
&  L^{\theta}_{\ot}  =  \delta  \CVA_{\ot}^{\theta} 
%+ \LGD_{\theh} 
- (\delta Z_{\ot}  )^\top\Delta 
 - c
 %\b{c/t=\E(-\delta { \rm HVA}_{\ot})/t
\eeql  
(cf.\ \eqref{e:Zt} and \eqref{e:pl_cva_sensi}).
As in Section \ref{ss:riskdyn}, the ``HVA trend'' $c$ (here ``$=-\E\delta {\rm HVA}_{\ot}={\rm HVA}_0-\E  {\rm HVA}_{\ot}$'') is deduced from $\Delta$ 
through the constraint that ${\mathbb{E}} L^{\theta}_{\ot}=0 $  (or $\widehat{\mathbb{E}} L^{\theta}_{\ot}=0 $ in the numerics).
By \textbf{EC and \PLE run-on sensitivities}, we mean
\begin{equation}\label{eq:ESMin}
     \Delta^{ec} = \argmin\limits_{ \Delta  \in \mathbb{R}^q } \ES\left(L^{\theta}_{\ot}\right)\mbox{ and }
     \Delta^{ple}  
    = \argmin\limits_{ \Delta \in \mathbb{R}^q
    }  \mathbb{E}  [  (%\tilde 
    L^{\theta}_{\ot} %\b{-c}
    )^2] ,
\end{equation} 
where $\ES$ means 95\% expected shortfall as in Section \ref{ss:hoff}.
Once $\delta \CVA^\theta_{\ot} $ learned from simulated $\varrho_{\ot}$ and $\xi_{0,T}(\varrho_{(t)})-\xi_{0,T}(\rho_0) $ 
%in the sensis mode 
the way mentioned after \eqref{e:pl_cva_sensi}, 
these sensitivities are computed much like their run-off counterparts of Section \ref{ss:riskdyn}.
Even simpler (but still optimized) \textbf{\LS 
 (run-on) sensitivities} are obtained without prior learning of $\delta\CVA_{\ot}$, just  by regressing linearly $\xi_{0,T}(\varrho_{\ot}) - \xi_{0,T}(\rho_0)$ against 
 $\delta Z_{\ot}$ the way explained after \eqref{e:pl_cva_sensi}  (purely linear LS regression here as opposed to linear diagonal quadratic LS regression in Table \ref{t:sensi}, due to our hedging focus of this part). 
 These \LS sensitivities are thus obtained much like the linear bump sensitivities of Algorithm \ref{a:fast_sensi}, except for the scaling of the bumps that are used in the corresponding  $\varrho_{(t)}$, and the fact that these LS sensitivities are computed directly in the market variables, without Jacobian transformation. 
 % In particular, the corresponding covariance matrix can no longer by inverted analytically: compare \eqref{e:varrho} and Line 1 with e.g.~$\sigma = 1\%, 3\%, 5\%$ in Algorithm \ref{a:fast_sensi}).
The derivation of the LS, EC, and PLE run-on sensitivities is summarized in Algorithm \ref{alg:reg_sensi}. Their computation times are reported in Table \ref{tab:time}. The right panels in Figure \ref{fig:risk_hedge}  page \pageref{fig:risk_hedge} show the run-on CVA hedging performance of different candidate sensitivities. All the risk compression ratios decrease with the risk horizon $t$. All sensitivities reduce both the unexplained PnL and economic capital by at least 2.5 times for $t=0.01$ and 4.5 times for $t=0.1$ and 1. Since client defaults are skipped in the run-on mode, the efficiency of bump sensitivities hedges is understandable.
For each risk horizon and performance metric, the optimized sensitivities always have better results than (benchmark or smart) bump sensitivities. Among those, the PLE and LS sensitivities hedges display the highest risk compression ratios and the lowest HVA trend $c$. Unlike what we observed in the run-off mode, most sensitivities (except for EC sensitivities when $t=0.01$ or $0.1$) also compress the HVA trend $c$ compared to the unhedged case. 

{
\def\hat{}  
\begin{algorithm}[H] 
\small
\LinesNumbered
\SetAlgoLined
\SetKwInOut{Input}{input}\SetKwInOut{Output}{output}
\Input{A set of calibrated initial values and model parameters $\rho_0 = (y_0, \epsilon_0)$ (client default indicators $X_0$ all set to 0), 
%a stochastic model parameter diffusion, 
a time horizon $t$, a number of exposure paths $m$ with $n$ pricing time steps.
%$m$ randomized $\rho_j=(Y_t^j(y_0, \epsilon_{(t)}^j\b{;??\tilde\omega_j}),\epsilon_{(t)}^j)$
}

\Output{Estimated sensitivities $\hat{\Delta}$.} 
Simulate $m$ cash flows $\xi_{0,T}(\rho_0;\omega_j)$ as per line 6 of Algorithm \ref{alg:cva_learning} in the baseline mode

Simulate $m$ realizations $\epsilon_j$ of $\varepsilon_{(t)}$ as per line 2 of Algorithm \ref{alg:cva_learning} with $\sigma = 1\%\sqrt{t}$ and $y_j=Y^j_t(y_0, \epsilon_j; \tilde\omega_j
%(\tilde\omega_j)_{\b{|[0,t]}}
)$ as per line 6 of Algorithm \ref{alg:cva_learning}, for Brownian drivers $\tilde\omega_j$ drawn independently from the $\omega_j$, and set $\rho_j=(y_j,\epsilon_j)$

Simulate $m$ realizations $\xi_{0,T}(\rho_j ;\omega_j )$ as per line 6 of  Algorithm \ref{alg:cva_learning} in the risk mode (with common random numbers, meaning here the same $\omega_j$ as in line 1)

For each $\rho_j$, $j=0\,..\,m$, compute the corresponding market hedging instrument prices scenario $z_j$

\uIf{{\rm LS}}{
Regress linearly the $\xi_{0,T}(\rho_j ;\omega_j) - \xi_{0,T}(\rho_0;\omega_j)$ against the 
$z_j -z_0$ 
%\b{??(or $\big(z_j -z_0, (z_j -z_0) \odot (z_j -z_0)\big)$ for linear quadratic approximation)} 
by SVD. The obtained coefficients are the LS sensitivities.  %(with small $L_2$ regularization)
}\Else{
Train a neural network 
$ \delta\CVA^{\theta}_{\ot}  (y,\epsilon)$ 
to regress the $\xi_{0,T}(\rho_j ;\omega_j) - \xi_{0,T}(\rho_0;\omega_j)$ against the 
$\rho_j -\rho_0.$

\uIf{{\rm EC}}{
Solve the left-hand side in \eqref{eq:ESMin}
by Adam stochastic gradient descent. 
}\ElseIf{{\rm \PLE}}{
Solve  the right-hand side in \eqref{eq:ESMin}
%(or \eqref{eq:LS} if model shifts are also hedged) 
by SVD linear regression. 
}
}

    \caption{LS, EC and \PLE run-on  sensitivities.\label{alg:reg_sensi}}  
\end{algorithm}}%def
%\vspace*{-0.01cm}
\begin{table}[H]%[htbp!]
\resizebox{\textwidth}{!}{
\centering
\begin{tabular}{|cc||ccccc||ccc|}
\hline
\multicolumn{2}{|c||}{{[}\Smart{]} bump sensitivities} & \multicolumn{5}{c||}{Regression sensitivities} & \multicolumn{3}{c|}{\multirow{2}{*}{Speedup}} \\ \cline{1-7}
\multicolumn{1}{|c|}{\multirow{2}{*}{\begin{tabular}[c]{@{}c@{}}Model\\ sensis\end{tabular}}} & \multirow{2}{*}{\begin{tabular}[c]{@{}c@{}}Jacobian\\ transform\end{tabular}} & \multicolumn{1}{c|}{\multirow{2}{*}{\begin{tabular}[c]{@{}c@{}}MtM\\ simulation\end{tabular}}} & \multicolumn{1}{c|}{\multirow{2}{*}{\begin{tabular}[c]{@{}c@{}}\LS\end{tabular}}} & \multicolumn{1}{c|}{\multirow{2}{*}{\begin{tabular}[c]{@{}c@{}} $\delta\CVA_{\ot}^\theta$  \\ learning\end{tabular}}} & \multicolumn{2}{c||}{\PnL regression} & \multicolumn{3}{c|}{} \\ \cline{6-10} 
\multicolumn{1}{|c|}{} &  & \multicolumn{1}{c|}{} & \multicolumn{1}{c|}{} & \multicolumn{1}{c|}{} & \multicolumn{1}{c|}{EC} & PLE & \multicolumn{1}{c|}{\LS} & \multicolumn{1}{c|}{EC} & PLE \\ \hline
\multicolumn{1}{|c|}{\begin{tabular}[c]{@{}c@{}}12min48s \\ {[}8.5s{]}\end{tabular}} & 30s & \multicolumn{1}{c|}{27s} & \multicolumn{1}{c|}{1s} & \multicolumn{1}{c|}{6s} & \multicolumn{1}{c|}{31s} & 1s & \multicolumn{1}{c|}{\begin{tabular}[c]{@{}c@{}}28.5 \\ {[}1.4{]}\end{tabular}} & \multicolumn{1}{c|}{\begin{tabular}[c]{@{}c@{}} 12.5\\ {[}0.6{]}\end{tabular}} & \begin{tabular}[c]{@{}c@{}}23.5 \\ {[}1.1{]}\end{tabular} \\ \hline
\end{tabular}
}
\caption{Computation times for learning $\delta\CVA_{\ot}^\theta$ and the related sensitivities. The speedups measure the ratios of the total time taken by the benchmark bump sensitivities approach to the total time taken by each of the sensitivities.}
\label{tab:time}
\end{table}

%\b{see possible to continue para} 

\section{Conclusion 
\label{s:concl}}
Table \ref{t:concl} synthesizes our findings regarding CVA (or more general, regarding columns 1 to 3) sensitivities, as far as their approximation quality to corresponding partial derivatives (for bump sensitivities) and their hedging abilities (regarding also the optimized sensitivities) are concerned. 
The winner that emerges as the best %(and always a good) 
trade-off for each downstream task in %bold color (
\textcolor{blue}{\bf blue}  
\textcolor{green}{\bf green} \textcolor{red}{\bf red}
%) 
in the first row is identified by the same color in the list of sensitivities. {\bf bench.\ bump} plays the role of market standard. 
Sensitivities that are novelties of this work are emphasized in \colorbox{yellow}{yellow} (\smart bump sensitivities essentially mean standard bump sensitivities with less paths, but with the important implementation caveat mentioned at the end of Section \ref{s:fast_ss}; PLE sensitivities were already introduced in the different context of SIMM computations in \citet{albanese2017var}; EC sensitivities were introduced in \citet{rockafellar2000optimization} and are also considered in \citeN{Buehler2017StatisticalH}).

% Sensitivities that are novelties of this work are emphasized in \colorbox{yellow}{yellow} (\b{including \smart bump sensitivities, even if these mainly mean standard \pointwise sensitivities with less paths, but with the implementation caveat mentioned at the end of Section \ref{s:fast_ss}}; PLE sensitivities were already introduced in the different context of SIMM computations in \citeN{albanese2017var}; EC sensitivities were introduced in \citeN{rockafellar2000optimization} and are also considered in \citeN{Buehler2017StatisticalH}).

\newcommand{\colorLocAcc}[1]{{\bf \color{blue} #1}}
\newcommand{\colorRunOn}[1]{{\bf \color{red} #1}}
\newcommand{\colorRunOff}[1]{{\bf \color{green} #1}}%teal
\newcommand{\cellnew}{\cellcolor{yellow}}
\begin{table}[H]
\resizebox{\textwidth}{!}{
\begin{tabular}{|l|l|c|c|c|c|c|}
\hline
%\multicolumn{2}{|c|}{sensitivities} 
& sensitivities & speed & stability & \colorLocAcc{local accuracy} & \begin{tabular}[c]{@{}c@{}}\colorRunOff{CVA run-off} \\ \colorRunOff{hedge}\end{tabular} & \begin{tabular}[c]{@{}c@{}}\colorRunOn{CVA run-on} \\ \colorRunOn{hedge}\end{tabular} \\ \hline
 & \textbf{bench.\ bump} & very slow & very stable & benchmark & bad & good \\ \hline
\multicolumn{1}{|l|}{\multirow{4}{*}{\begin{tabular}[c]{@{}l@{}}fast \\ bump\\ sensis\end{tabular}}} & naive AAD bump & fast & fragile & \textit{bad} & \textit{bad} & \textit{bad} \\
\multicolumn{1}{|l|}{} & \cellnew AAD bump & fast & fragile & \textit{average} & \textit{bad} & \textit{bad} \\
\multicolumn{1}{|l|}{} & \cellnew linear bump &  fast & average & good & \textit{bad} & \textit{good} \\ 
\multicolumn{1}{|l|}{} &  \colorLocAcc{\smart bump} &  fast & stable & good & \textit{bad} & good \\ \hline
\multicolumn{1}{|l|}{\multirow{3}{*}{\begin{tabular}[c]{@{}l@{}}optimized\\ sensis\end{tabular}}} & EC sensis & fast & average & not applicable & good & very good \\
\multicolumn{1}{|l|}{} & \colorRunOff{PLE sensis} & very fast & stable & not applicable & good & excellent \\
\multicolumn{1}{|l|}{} & \cellnew \colorRunOn{LS w/o $\boldsymbol{\Gamma}$} & very fast & stable & not applicable & not applicable & excellent \\ \hline
\end{tabular}}
\caption{Conclusions regarding sensitivities and hedging.
% as far as their approximation quality to corresponding partial derivatives (for bump sensitivities) and their CVA hedging abilities (regarding also the optimized sensitivities) is concerned. 
By local accuracy of a bump sensitivity, we mean the accuracy of the approximation it provides for the corresponding partial derivative. {\em Italics} means tested but not reported in tables or figures in the paper.}
\label{t:concl}
\end{table}
\vspace*{-0.35cm}
Regarding the assessment of CVA risk, in the run-on CVA case (see Table \ref{t:conclrisk} using the same color code as  Table \ref{t:concl}), we found out that neural net regression of conditional CVA results in likely more reliable (judging by the twin scores of the associated CVA learners) and also
faster value-at-risk and expected shortfall estimates than CVA Taylor expansions based on bump sensitivities (such as the ones that inspire certain regulatory CVA capital charge formulas). But an LS proxy, linear diagonal quadratic in market bumps, provides an even quicker (as it is regressed without training) and almost equally reliable view on CVA risk as the neural net CVA. In the \textbf{\b{run-off CVA}} case (not represented in the table), the \textbf{\b{neural net learner of CVA$_t$}} in the risk mode (or nested CVA learner alike but in much longer time) allows one to get a consistent and dynamic view on CVA and counterparty default risk altogether. 
\begin{table}[H]
\resizebox{\textwidth}{!}{
\begin{tabular}{|l|c|c|c|c|c|}
\hline
 & $\delta \CVA_{\ot}$ learners & speed & stability & \colorLocAcc{twin accuracy} & $\delta \CVA_{\ot}$ VaR and ES \\ \hline
\multirow{2}{*}{nonparametric} & \cellnew nested MC & very slow & stable & very good & very conservative \\
 & \cellnew neural net & fast & average & good & conservative \\ \hline
\multirow{2}{*}{\begin{tabular}[c]{@{}l@{}}linear(-quadratic)\\ in market bumps\end{tabular}} & \textbf{bench.\ bump} & very slow & very stable & \begin{tabular}[c]{@{}c@{}}good/average\\ for small/large $t$\end{tabular} & aggressive \\
 & \cellnew \colorLocAcc{LS w/ $\boldsymbol{\Gamma}$ in $\boldsymbol{\delta Z_t}$} & very fast & stable & good & conservative \\ \hline
\end{tabular}
}\caption{Conclusions regarding run-on CVA risk.}\label{t:conclrisk}
\end{table}

% This paper is a contribution to CVA along three lines :
% (i) for accelerating via our so called linear bump sensitivities the computation of the CVA bump sensitivities needed for hedging purposes and \b{SA CVA VaR} calculations;
% % (but we don't advocate the latter for internal model risk)
% (ii) for devising optimized CVA hedging schemes  
% regarding a CVA assessed either on  a run-off or a run-on basis;
% (iii) for internal  modeling purposes related to the assessment of provisions for CVA model risk (in the form of a related hedging valuation adjustment) and of economic capital for CVA and/or counterparty default risk. 

% \b{In particular, if one wants to account for defaults in CVA computations and hedging schemes on a run-off portfolio and customers  basis, market \pointwise bump sensitivities may be too local. Alternative market sensitivities suitably optimized for this purpose are studied in Section \ref{ss:hoff}.
% A nonlocal variation on our linear bump sensitivities, namely the \LS sensitivities of Section \ref{s:cpes} (see Algorithm \ref{alg:reg_sensi} page \pageref{alg:reg_sensi}), also yield a \smart and good proxy for CVA risk assessed on a run-on portfolio and customers basis}.

\bibliographystyle{chicago}
\bibliography{./manuscript}
 
\end{document}